\documentstyle[preprint,aps,epsfig]{revtex}

\newcommand{\nc}{\newcommand}
\def\frac#1#2{{\textstyle {#1 \over #2}}}

\nc{\beq}{\begin{equation}}
\nc{\eeq}{\end{equation}}
\nc{\beqa}{\begin{eqnarray}}
\nc{\eeqa}{\end{eqnarray}}
\nc{\lsim}{\begin{array}{c}\,\sim\vspace{-21pt}\\< \end{array}}
\nc{\gsim}{\begin{array}{c}\sim\vspace{-21pt}\\> \end{array}}
\nc{\p}{\partial}
\nc{\nn}{\nonumber}

\begin{document}
\draft
\title{
A Monte Carlo analysis of the phase transitions in the 
$2D$, $J_1-J_2$ XY model
}
\author{D. Loison$^a$ and  P. Simon$^b$}
\address{
$^a${\it  Institut f\"ur Theoretische Physik, Freie Universit\"at Berlin,
Arnimallee 14, 14195 Berlin, Germany\\ Damien.Loison@physik.fu-berlin.de}\\
$^b${\it International School for Advanced Studies, Via Beirut 2-4,
34014 Trieste, Italy\\simon@he.sissa.it}
}
\maketitle

\begin{abstract}
We consider the $2D$ $J_1-J_2$ classical XY model on a square lattice.
In the frustrated phase corresponding to $J_2>J_1/2$, an Ising like
order
parameter emerges by an ``order due to disorder'' effect. This leads to
a
discrete $Z_2$ symmetry plus the $U(1)$ global one.
Using a powerful algorithm we show that the system undergoes two
transitions
at different but still very close temperatures, one of
Kosterlitz-Thouless
(KT) type and another one which does
not belong to the expected Ising universality class. A new analysis of
the KT
transition has been developed in order to avoid the use
of  the non-universal helicity jump and to allow the computation of the
exponents without a precise determination of the critical temperature.
Moreover,
our huge number of data enables us to exhibit the existence of large
finite
size effects explaining the dispersed results found in the literature
concerning the  more studied  frustrated $2D$, XY models.
\end{abstract}
\vspace{1.cm}
PACS NUMBERS: 05.50.+q, 75.10.Hk, 05.70.Fh, 64.60.Cn, 75.10.-b

\section{Introduction}

The ground state of a large class of two-dimensional classical frustrated XY
models have the particularity to exhibit both continuous and discrete $Z_2$
degeneracy simultaneously in the ground state. It results   the appearance of
a new Ising-like order parameter, in addition to the continuous $U(1)$ symmetry.
The most famous example exhibiting such behaviors is certainly the fully
frustrated XY model (FFXY) which was originally introduced by Villain           
as a frustrated XY model without disorder \cite{vil1}. In this model,
the $Z_2$ symmetry is associated with the two types of chirality ordering.
This model is also of great interest because it describes a superconducting
array of Josephson junctions under an external transverse magnetic field such
that the flux per plaquette is half the quantum flux \cite{joseph}.
For fifteen years, extensive (essentially numerical) works have been carried
on the FFXY \cite{diep}-\cite{benakli}
and also to some related models like the triangular antiferromagnetic XY
model\cite{trian}, the helical XY model\cite{heli}, the Coulomb gas system
of half integer charges \cite{olsson,jrlee}. The interplay between
the two transitions can lead {\it a priori} to two transitions, namely a
Kosterlitz-Thouless one and an Ising one. Nevertheless, the entanglement
between the two order parameters considerably complicates the analysis.
The nature of the phase diagram is still rather unconclusive and controversial. 
Three different scenari have been advocated:  either  the two transitions
occur at the same point and eventually merge to give a new universality class
\cite{kos1,jose,knops,granato};  or  the two transitions occur at different
points and are of Ising and Kosterlitz-Thouless types plus some strong finite
size effects \cite{olsson,koslee}; or finally  the two transitions are
effectively separated but  the transition
 associated to the chiral order parameter is not of Ising type
\cite{slee,luo,jrlee}. 
The most recent numerical studies are in favor of the latter scenario.
Nonetheless, without a strong analytical support, the problem is still
completely open.

The purpose of the present article is other. We want to clarify the critical
behavior of a less studied frustrated XY model: the $2D$, $J_1-J_2$ XY
model on a
square lattice which, as will be shown, is in the same universality class 
as the models quoted above (or more precisely has the same problematic). 
The  Hamiltonian reads
\begin{eqnarray}
\label{ham}
H&=& -J_1\sum\limits_{<i,j>}{\bf S}_i{\bf S}_j+J_2\sum\limits_{<<k,l>>}
{\bf S}_k{\bf S}_l \\
&=&-J_1\sum\limits_{<i,j>}\cos(\theta_i-\theta_j)+J_2\sum\limits_{<<k,l>>}
\cos(\theta_k-\theta_l)~,
\end{eqnarray}
\noindent
where ${\bf S}_{i}$ are two component classical vectors of unit length,
with $J_1,J_2>0$, $<~>$ and $<<~>>$ indicate respectively  the sum over
nearest neighbors and next to nearest neighbors.  When $2J_1>J_2$ the
ground state is ferromagnetic. It leads to a Kosterlitz-Thouless (KT)
transition at the temperature $T_{KT}\approx {\pi(J_1-2J_2)\over 2}$ \cite{sim}.
This temperature is obtained from the Villain treatment of the Hamiltonian (\ref{ham})
(see ref. \cite{sim1} for details).
However, when $2J_1<J_2$, the ground state consists of two independent
$\sqrt{2}\times\sqrt{2}$ sublattices with AF order. The ground state
energy $E_0=-2NJ_2$ does not depend on $\phi$, an angle parameterizing
the relative orientations between  both sublattices. This non-trivial
degeneracy
is lifted by thermal fluctuations and a collinear ordering (corresponding
to $\phi=0$ or $\pi$) is selected \cite{hen}. The two possible ground states 
are depicted in Fig.~\ref{gs}. The angle $\phi$ thus
plays a role analogous to the chiral  order parameter. This selection
mechanism is one of the most famous ``order due to disorder'' effect
\cite{hen} in a sense that fluctuations brings kind of order by lifting
this extra continuous symmetry. 
The resulting symmetry is therefore $U(1)\times Z_2$.
Monte-Carlo simulations predict only a low temperature phase with a
nematic ordering and a disordered  high temperature phase  \cite{hen,fernan}.
The critical behavior has, as far as we know, only been partially explored in
ref. \cite {fernan}. 
Unfortunately the results are very approximate and no definite conclusion on
the presence of one or two transitions and either on their universality
classes has been given.
In this work, we have carried on extensive numerical Monte Carlo simulations
on the $J_1-J_2$ XY model using algorithms which allow us to obtain very 
accurate and robust  results.

We now give the outline of the paper. Section 2 contains a brief summary
of analytical results showing the relations between the $J_1-J_2$ XY model
and the
Ising-XY model, which is a generic model used to describe the universality
class of frustrated XY models with symmetry $Z_2\times U(1)$ like the FFXY.
In section 3, we present our numerical results and the analysis of some
critical exponents. Finally, section 4 is devoted to the discussion of the
results and to a brief conclusion.
The estimation of statistical and systematic errors has been relegated 
in the appendices.  

\section{The $J_1-J_2$ XY and Ising-XY model}

In this section, we sum up the main analytical results concerning the
$J_1-J_2$ XY model. We essentially focus on the more interesting
frustrated phase 
corresponding to $2J_2>J_1$, where the ground state consists of two
independent sublattices. Thermal fluctuations select a collinear
ordering \cite{hen}, and we have two kinds of domains represented in
Fig.~\ref{gs}.
The first step, following Chandra et al. \cite{chan}, is to perform a
gradient expansion of the classical energy (\ref{ham}). The problem is
now translated in a new one on a $(2\times 2)$ square lattice, but now
with two spins $1$ and $2$ per vertices pointing in the same directions.
The new classical action ${\cal A}$ reads
\begin{eqnarray}
\label{action}
{\cal A} &=&-{2J_2\over 2T}\sum\limits_r \left[
(\vec{\nabla}\theta_1)^2+ (\vec{\nabla}\theta_2)^2\right.\nn\\ 
&+& \left. 2\lambda\cos\phi~~ (\nabla^x\theta_1\nabla^x\theta_2
-\nabla^y\theta_1\nabla^y\theta_2)\right] ~,
\end{eqnarray}
\noindent
where we have defined $\lambda={J_1\over 2J_2}<1$ and introduced the
lattice derivatives $\nabla^x,~\nabla^y$ \cite{sim}. The signature of
the $U(1)$ degeneracy lies now in the strong anisotropy between $x$
and $y$ directions. The $\cos\phi$ labels the different possible
classical ground states at $T=0$.
Notice that, if we do the Gaussian integration over the angular variables,
we recover the results of Henley \cite{hen}, namely
\begin{eqnarray}
\label{asw}
{\cal A}\sim const-0.32\left({J_1\cos\phi\over 2J_2}\right)^2~,
\end{eqnarray}
proving that a collinear ordering is selected when minimizing the free
energy($\cos(\phi)=\pm 1$).

Let us now include the effects of the vortices. The most natural way to
include them would be to apply the Villain transformation to all quadratic
terms in the action (\ref{action}). Such a treatment is quite tedious and
unappropriated because the vortices variables built
with the anisotropic term
$(\nabla^x\theta_1\nabla^x\theta_2-\nabla^y\theta_1\nabla^y\theta_2)$
are not well defined due to the strong fluctuations between the two sublattices
which  tend to decouple in the infrared limit ( see \cite{sim} for
more details). A simplest way to
take into account the coupling between the two sublattices is to
replace the anisotropic term in (\ref{action}) by the local spin waves
effect {\it i.e.}
by $-0.32 \lambda^2\sum\limits_r \cos^2(\theta_1(r)-\theta_2(r))$. This
term is just the local version of (\ref{asw}).
Such a treatment has already been used by Garel {\it et al.} for helimagnets
\cite{heli} and also by Chandra {\it et al. } for the $J_1-J_2$ Heisenberg
model \cite{chan}.
By applying the Villain transformation to the first two terms in
(\ref{action}) and using
\begin{eqnarray}
\exp\left[h\cos p (\theta_1(\vec{r})-\theta_2(\vec{r}))\right]
=&&\sum\limits_{S(\vec{r})}\exp \left[ipS(\vec{r})
(\theta_1(\vec{r})-\theta_2(\vec{r}))\right.\nn\\&&+\left.\log y_sS^2(\vec{r})\right],
\end{eqnarray}
\noindent
with  $p=2$ and $y_s=h/2=0.08\lambda^2$, we obtain

\begin{eqnarray}
{\cal Z}= \sum\limits_{\{n_1^{\mu}(r)\}}
 \sum\limits_{\{n_2^{\mu}(r)\}} \sum\limits_{S(r)}&&\int {\cal D}\theta_1
{\cal D}\theta_2
\exp \left( -{J_2\over T}\sum\limits_r\sum\limits_{i=1,2} 
\left[ (\nabla^{\mu} \theta_i(r)-2\pi n_i^{\mu}(r))^2
\right]\right.\nn\\
&+&ip\left.\sum\limits_r\left[ S(r)(\theta_1(\vec{r})-\theta_2(\vec{r}))
+\log y_sS^2(\vec{r})\right]\right)
\end{eqnarray}
The $n_i^{\mu}$ ($i=1,2$) are integer link variables living on the two
diagonal sublattices. By integrating on angular variables, we easily find
\begin{eqnarray}
\label{actfin}
{\cal A}_{eff}&&=\sum\limits_{r\neq r'}\left[\pi\beta  J_2 M_1(r)
\log{|r-r'|\over a} M_1(r')
+\pi\beta J_2 M_2(r) \log{|r-r'|\over a} M_2(r')\right.\nonumber\\
&&-ip (M_1(r)+M_2(r))  \Theta|r-r'| S(r')
+\left.{p^2\over 2\pi\beta} S(r)\log {|r-r'|\over a}S(r')
\right]\\&&+\sum\limits_r\left[ \log y_1~ (M_1)^2(r)+\log y_2~ (M_2)^2(r)
+\log y_s~ S^2(r)\right]~\nonumber,
\end{eqnarray}
where we have introduced the vortex variables 
$M_i=\epsilon^{\mu\nu}\nabla^{\mu}n_i^{\nu}$ corresponding to vortices on
the two sublattices. The fugacities are as usual considered as genuine
variables defined initially by $y_i^0=\exp(-{\pi^2\beta J_2\over 2})$.
The interaction $\Theta$ is defined by
$\Theta |r-r'|=\arctan({y-y'\over x-x'})$, and verifies
$\p_y\log|r-r'|=-\p_x \Theta|r-r'|$. This action corresponds to two
coupled XY models.
Under real space renormalization group transformations, the coupling term is 
strongly relevant and locks the phase difference in
$\theta_1(r)=\theta_2(r)+k\pi$ with $k=0,1$ \cite{doma}.
It leads in the strong coupling limit to an 
effective  model whose Hamiltonian has the following form

\begin{eqnarray}
\label{isxy}
H_{I-XY}=-\sum\limits_{<i,j>}[A(1+\sigma_i\sigma_j)\cos(\theta_i-\theta_j)
+C \sigma_i\sigma_j]~.
\end{eqnarray}
\noindent
The value of $A$ and $C$ depend of the initial values of $h=f(J_1,J_2)$
and $\beta J_2$. This model refereed as the Ising-XY model in the literature
has been largely debated. This Ising-XY model is believed to describe the
critical behavior of all frustrated XY models quoted in the introduction.
The phase diagram has been obtained numerically
by Granato {\it et al.} \cite{isxy} and has been reproduced
for convenience in 
Fig.~\ref{fig.I-XY}. 
Three different phases can be distinguished: the upper right corner phase
correspond to the low temperature ordered phase, the low left corner phase
is the high temperature disordered one, and the intermediate one is Ising
ordered but XY disordered (namely with free vortices).
Above the point $P$, the Ising
and XY transitions are well separated and mix under P. The question
concerning the transition(s) under $P$ is still under debate. A recent work
of S. Lee {\it et al.}
seems to indicate that the two transitions never merge completely but get
closer \cite{koslee}. Nevertheless the Ising-like magnetization exponent has
been found different from $1$ and continuously varying along the line (PT)
\cite{isxy}. 
We have shown that the $J_1-J_2$ XY model should be also described by the
Ising-XY model and should therefore correspond to  a curve crossing the
line under P (so with only one  or two very close transitions).  Since we
are not able
to provide analytical relations between $(h,J_2)$ and $(A,C)$, the form of
this curve and its intersection with the segment (PT) is unaccessible.
Moreover,
when varying $J_2/J_1$, we shall obtain a different intersection point as
it was firstly noted in \cite{sim}. 
Nevertheless, it opens the possibility that the critical exponent $\nu$
should vary with the ratio $J_2/J_1$ as in the analysis of Granato {\it al. }
\cite{isxy} or of Lee {\it al.} \cite{koslee}. 
Similar considerations have been done in the study of a generalized
frustrated XY model where an extra-parameter has been added \cite{benakli}.
No clear conclusion concerning the nature of the phase transitions
can be therefore  drawn at this level. The purpose of the next
section  is therefore to answer these questions by help of  extensive
Monte Carlo simulations. Moreover, it can also be regarded  as an indirect
way of studying the Ising-XY model and other related models.

\section{Monte Carlo analysis}
\subsection{Observables}
As explained above we can define two order parameters corresponding to the two
 symmetries $U(1)$ and $Z_2$. The first one is the total magnetization $M$
defined
by the sum of all  spins, the second is the chirality $\kappa$ defined by
the sum of all chiralities $\kappa_i$ defined on each cell by:
\begin{eqnarray}
\label{formulekappa}
\kappa_i={1 \over 4}\ (S_i-S_k)(S_j-S_l) 
\end{eqnarray}
where ($i,j,k,l$) are the four corners of one cell with diagonal (i,k) 
and (j,l). The two ground states depicted in Fig.~\ref{gs} have
$\kappa_i=\pm1$. 

We have studied our system in the finite size scaling region where the 
correlation length is much bigger than the lattice size.  
The quantities needed for our analysis are defined below.
For each temperature we calculate:
\begin{eqnarray}
\label{formuleXM}
\chi^M &=& {{N<M^2>} \over {k_{B}T}} \\
\label{formuleXK}
\chi^{\kappa} &=&{ {N(<\kappa^{2}>-<\kappa>^2)}\over {k_{B}T}} \\
\label{formuleXK2}
\chi_2^{\kappa} &=& {{N<\kappa^{2}>}\over {k_{B}T}} \\
\label{formuleV1}
V_{1}^\kappa&=& {{<\kappa E>}\over {<\kappa>}}-<E> \\
\label{formuleV2}
V_{2}^\kappa&=&{ {<\kappa ^2E>}\over {<\kappa^2>}}-<E> \\
\label{formuleV2.M}
V_{2}^M&=& {{<M ^2E>}\over {<M^2>}}-<E> \\
\label{formuleUM}
U^M&=&1-{ {<M^{4}>}\over {3<M^{2}>^{2}}}\\
\label{formuleUK}
U^{\kappa}&=&1-{{<\kappa^4>}\over {3<\kappa^2>^2}}
\end{eqnarray}
where
$E$ is the energy,  
$\chi$ is the magnetic susceptibility per site, 
$V_{1,2}$ are cumulants used to obtain the critical exponent $\nu$,
$U$ are the fourth order cumulants,
$<...>$ means the thermal average.

\subsection{Algorithm}
We use in this work the standard Metropolis algorithm. 
At each site  a new random orientation for the spin is chosen. 
The interaction energy between this spin and its neighbors is then calculated.
If  lower than the energy of the old state, the new state is
accepted,  otherwise, it is accepted only with a probability $p$ according to
the standard Metropolis algorithm.

However the critical slowing down is important and we  improve the speed
of the simulation using the local over-relaxation algorithm (OR) 
\cite{Overrelaxation}. This algorithm consists in  choosing the new orientation
of the spin such that the energy remains unchanged. For $XY$ spins 
the only possibility is to take the symmetric of the old spin 
to the local field (the sum of the neighbors). This algorithm is obviously
non ergodic, i.e. all states can not be reached. It must thus  be used in 
combination with the standard Metropolis algorithm (MET). Therefore,
at each step
(regarded as one unit) we use one MET step  and  a certain  number of steps
of over-relaxation ($NOR$) algorithm.  
The larger $NOR$, the  smaller the autocorrelation time  
(the number of steps
between two independent configurations), but then the  larger the time
needed for
each step is. We have thus to choose the best  $NOR$ to minimize the real 
autocorrelation time. This depends on the
time needed for each algorithm. In our implementation the Metropolis algorithm 
is six times larger than the over relaxation algorithm. 

In order to calculate the autocorrelation time we follow the procedure explained
in appendix \ref{annexe1}.
In table \ref{table1} we present the results of the
autocorrelation time $\tau_\kappa$ for different $NOR$ at 
the critical temperature $T_c^\kappa$ for a lattice size $L=30$ and
$J_2/J_1=0.7$.
The second column gives $\tau_\kappa$ while the third column gives the real
autocorrelation time $\tau_\kappa \, (1+NOR/6)$, i.e. the quantity to be 
minimized.
The value $NOR\sim L/5=6$ seems to fit best. 
We have checked that this ratio does not change significantly for sizes $L=20$
and $L=40$, which is in
agreement with the argument of Adler \cite{Adler81} where the best $NOR$ should
be proportional to the correlation length, i.e. to the lattice size in the 
finite size scaling region.
 
In Fig~\ref{figuretau} we have shown in a log-log scale 
the real chirality autocorrelation time 
 function of the lattice size for $NOR=0$ and $NOR=L/5$. 
For larger lattice sizes the gain is more than a factor 30 using the over 
relaxation algorithm.
The critical exponent $z$ defined by $\tau\sim L^z$ is 2.29(4) 
without the use of the OR 
algorithm and is in agreement with the results on the Villain lattice 
 2.31 \cite{Pawig98} but in disagreement with the dynamic approach of Luo et al.
\cite{luo} who obtained 2.17(4). We note for this last case that an error
on $z$ leads to errors on the other exponents.

In the following the simulations have been done using $NOR=L/5$  for each
Metropolis algorithm.
For each simulation, we use a number $t_{av}$ measurements, made after 
an updating time $t_{up}$ is carried out for equilibration. For each size,
between 
three and six different initial configurations (ordered or random) have been
tested to be sure that our system is not trapped into a metastable state. 
In table \ref{table2} we present some details of our simulations. 
We want to 
stress that the number of Monte Carlo steps used in this work is one order
of magnitude larger 
than  previous studies and, combined with a better algorithm, produces
a better estimate of the thermodynamic quantities.

Our errors are calculated with the help of the Jackknife procedure
\cite{Jackknife}. When compiling the different results from previous
studies 
we have noticed that the errors reported are quite often strongly 
underestimated. Therefore  we have presented in the appendix \ref{annexe1}
our method to evaluate the errors coming from the simulation and in particular
a simplified method of the Jackknife procedure. 

We use in this work the histogram MC technique
developed by Ferrenberg and Swendsen \cite{Ferren88,Ferren89}.
From a simulation done at $T_{0}$,
this technique allows to obtain thermodynamic quantities
at $T$ close to $T_{0}$.

\subsection{Phase diagram}
We have performed many simulations in varying the value of $J_2/J_1$ to
obtain the critical temperature $T_c$ 
which is represented in Fig~\ref{figureTc}.
The transition for $J_2/J_1<0.5$ is a standard
Kosterlitz-Thouless transition in agreement with theoretical predictions.
For $J_2/J_1>0.5$ it is difficult to
discriminate between the hypothesis of one or two transitions separating
the low temperature nematic phase from the high temperature disordered phase.
We have therefore decided to focus on the particular value $J_2=0.7,J_1=1.$
(black circle) in the remainder of this work. It is worth noticing
that  Fernandez et al. \cite{fernan} 
have done their calculation for $J_2=J_1=1$.
    
As it was first emphasized in section 2, it is possible that the exponents could vary 
with the ratio $J_2/J_1$ \cite{sim}. This should be coherent with the picture proposed
by Granato et al for the Ising-$XY$ model \cite{isxy} and by Lee {\it et al.} \cite{koslee}.
Nevertheless, the first step is to perform very highly accurate Monte Carlo
simulations at some fixed value of $J_2/J_1$ to show the existence
to test the existence of two close transitions, and check that the chiral 
magnetic exponent $\nu$ is clearly different from $1$.

\subsection{$Z_2$ symmetry}
We concentrate first on the breakdown of the $Z_2$ symmetry with the 
order parameter $\kappa$ defined in (\ref{formulekappa}).

To find the critical temperature $T_{c}$ we record the variation of $U^\kappa$
with $T$ for various system sizes in Fig~\ref{figCM}
and then locate $T^\kappa_c$ at the intersection
of these curves \cite{BinderU} since
the ratio of $U^\kappa$ for two different lattice sizes $L$
and $L'=bL$ should be 1 at $T^\kappa_c$, namely
\begin{equation}
\label{BinderU}
{{{U^\kappa_{bL}}\over {U^\kappa_{L}}}}\Bigg\arrowvert _{T=T_{c}} = \> 1 \> .
\end{equation}
Due to the presence of residual corrections to finite size scaling,
one has actually to extrapolate the results taking the limit
(ln$b$)$^{-1} \rightarrow  0$ in the upper part of the Fig.~\ref{figTc2.all}. 
We observe a strong correction for the small sizes. However for the biggest
sizes the fit seems good enough and we can
extrapolate $T^\kappa_c$ as
\begin{equation}
\label{tgtg}
T_c^\kappa= 0.56465(8)\ ,
\end{equation}
The estimate for the universal quantity $U^\kappa_*$ at the
critical temperature is
\begin{equation}
U^\kappa_*=0.6269(7).
\end{equation}
This value is far away of the two dimensional Ising value 
$U_*^{Ising}\sim0.611$,
\cite{Kamierenarz-Blote} which is a strong indication that the universality
class associated to  the chirality order  parameter is not of Ising type. We
will  verify this prediction studying now the critical exponents. 

At $T=T^\kappa_c$ the critical exponents can be determined by log--log fits. 
We obtain $\nu^\kappa$ from $V_1^\kappa$ and $V_2^\kappa$ (Fig.~\ref{figV}),
$\gamma^\kappa/\nu^\kappa$
from $\chi^\kappa$ and $\chi_2^\kappa$ (Fig.~\ref{figX.all}),
and $\beta^\kappa/\nu^\kappa$ from $\kappa$ (see Fig.~\ref{figM}).
We observe in all these figures a strong correction to a 
direct power law. It is worth noticing  however that $X_2^\kappa$ shows 
smaller corrections.
Using only the three (four for $X_2^\kappa$) largest terms
we obtain: 
\begin{eqnarray}
\label{formule.nu.K}
\nu^\kappa&=&0.795(20)\\
\label{formule.gammasnu.K}
\gamma^\kappa/\nu^\kappa &=& 1.750(10) \\
\label{formule.betasnu.K}
\beta^\kappa/\nu^\kappa &=& 0.127(10)\ .
\end{eqnarray}
The uncertainty of $T^\kappa_c$ is included in the estimation of the errors.
The large values in our errors come from the use of only few sizes 
for our fits. If we had used more, the exponents would change and for
example $\nu^\kappa$ 
would grow until 0.91 using all the sizes.
The non observation of the corrections 
in previous studies could explain 
the very dispersed results obtained in various studies of different frustrated
$2D$ XY models (between 0.76 to 0.90). 
We note that we have used much more statistics (due to one part to a 
better algorithm, and in other part to longer simulations) than previous
works (between one or two order of magnitude more) which enables us to observe 
the finite size corrections.
Moreover, we expect that the critical exponents written above could vary with 
the ratio $J_2/J_1$. It makes therefore difficult quantitative comparisons
with other studies. Nevertheless, we can safely state that an Ising universality class is excluded.
If we try to introduce a correction to 
calculate the
exponents, for example like  $V_1^\kappa=(1+L^{-\omega})L^{1/\nu^\kappa}$,
we  obtain
$\omega=1.0(2)$ and values for critical exponents fully compatible
with (\ref{formule.nu.K}-\ref{formule.betasnu.K}). 
We have noticed that the exponents have a tendency
to move away from the ferromagnetic Ising values when the size grows and thus 
seems to exclude a crossover to the ferromagnetic Ising fixed point for
larger sizes (unless it occurs at very large and not yet accessible size). 

The values given in (\ref{formule.nu.K}-\ref{formule.betasnu.K}) use the
properties of the free energy at the critical temperature. But an error
on $T^\kappa_c$ leads to an error on the exponents,  
it is therefore interesting to
find them without the help of $T^\kappa_c$. This can be done using 
the whole finite size scaling region and the method given in 
\cite{Loison.O6.Ferro}. It consists to plot, for example, the susceptibility 
$X^\kappa L^{-\gamma^\kappa/\nu^\kappa}$ as function of $U^\kappa$, choosing
the exponents 
as the curves collapse. 
This fit is stronger than the fit at the
critical temperature in so far as it does not depend only on results at
$T^\kappa_c$ but on a large region of temperature. 
However the errors are a little bit larger. We show in 
Fig.~\ref{X2.1}-\ref{X2.3} the results for three choices of 
$\gamma^\kappa/\nu^\kappa$. 
Obviously the result 
$\gamma^\kappa/\nu^\kappa=1.76$ is the best one. With this method we 
arrive at 
$\gamma^\kappa/\nu^\kappa =1.76(2)$
which is compatible with the result (\ref{formule.gammasnu.K}) and thus
constitutes  an indirect 
check of the critical temperature. We have verified,
using cumulants $V_1$ and $V_2$ and $<\kappa>$, that the results for 
$\nu^\kappa$ and
$\beta^\kappa/\nu^\kappa$ are compatible with (\ref{formule.nu.K}) and 
(\ref{formule.betasnu.K}).

From the scaling
relation
\begin{equation}
\label{hyperscaling}
\gamma^\kappa/\nu^\kappa = 2 -\eta^\kappa
\end{equation}
we obtain $\eta^\kappa=0.25(1)$.
The results are summarized in table \ref{table3}.

In conclusion the chirality order parameter does not seem to belong to the
standard two dimensional Ising universality class. Such conclusion has
already been reached in many studies of frustrated XY models.
Nevertheless, due to the fact that the exponents
could be $J_2/J_1$ dependent, we cannot safely compare the results we get
for the $J_1-J_2$ XY model with other frustrated XY models. 
However we observe that the exponents vary strongly if corrections
are not taken into account and we suspect that this is also the case in
the other studied models.

If the transition belongs to a new universality class, the use of
Binder's cumulant at the critical temperature $U^\kappa_*$ could be a new
approach to track it. It should be 
very interesting to test this idea in other systems 
like the Villain or the triangular models.

\subsection{$U(1)$ symmetry}
We  now focus on the phase transition associated to  the $U(1)$ 
symmetry, i.e. to  the $XY$ spins. 
The usual ferromagnetic XY model undergoes a Kosterlitz-Thouless  phase
transition  driven by the unbounding of 
vortex-antivortex pairs. The best method to obtain reliable results 
is to use the jump of the helicity parameter defined by the answer of the
system to a twist in one direction. The knowledge
of the jump at the critical temperature allows to obtain $T_c$ with a good 
precision \cite{Olson95}. However, in our problem the presence of the
chirality order 
parameter coupled with topological defects leads to a non-universal
jump. This fact explains why
this transition is usually not explored in Monte-Carlo simulations of
frustrated XY models or, when
it is, why results are not very accurate. In the following study we will
use a method introduced in \cite{LoisonO2Ferro} using Binder's cumulant
to study this transition. It was proved in this article that, contrary to
the common belief, the Binder cumulant for
ferromagnetic XY systems  
crosses for different sizes, allowing thus  a rough estimate of the 
critical temperature and especially of the exponent $\eta$ without the precise
knowledge of the critical temperature. 

We proceed in a way similar as for the Ising order parameter, replacing
$\kappa$ by $M$.
We record the variation of $U^M$ (\ref{formule.U.M}) with the temperature
for various system sizes in Fig~\ref{figCM.KT}. 
We want to underline the differences between the result of $U^\kappa$ 
(Fig~\ref{figCM.KT}) and $U^M$ which are plotted with the same scale. 
$U^M$ shows a crossing on a smaller region than $U^\kappa$, and  at least one
order of magnitude less than the standard XY model 
(see the figure 1 of \cite{LoisonO2Ferro}) 
We then locate the intersection of these curves and plot
the results in the lower part of the Fig.~\ref{figTc2.all}. 

Let us  first  consider a power law behavior at $T>Tc$ for this system. 
In this case we have 
to consider a linear fit for (ln$b$)$^{-1} \rightarrow 0$. We observe 
corrections for the smallest size $L=20$ but the others seem to converge
to the temperature 
\begin{equation}
\label{formule.Tc.M}
T_{c}^M= 0.56271(5)\ .
\end{equation}

Secondly we consider  the behavior to be exponential as in the standard 
$XY$ model. In this case 
figure 2 of \cite{LoisonO2Ferro} shows that a linear fit could be wrong
and that a "crossover" to a different critical temperature could be 
observed for bigger $b$, i.e. greater sizes. However contrary to 
the ferromagnetic XY model, the region of crossing is so small and the
different
linear fits tend only to one critical temperature.  Therefore we think that
the linear fit works well enough.  Moreover in the following we will show 
strong arguments in favor of the temperature (\ref{formule.Tc.M}).

With the help of the critical temperature we have found an estimate 
of $U^M$ at $T_c^M$ fitting the value with a law $U^M=U^M_*+aL^{-\theta}$.
We obtain
\begin{eqnarray}
\label{formule.U.M}
U^M_*=0.638(5)\ .
\end{eqnarray}
By log--log fit we calculate some exponents. The exponent $\eta$ could 
be obtained by a fit with
$X_2^M$ shown in Fig~\ref{figX.all}. We obtain here
\begin{eqnarray}
\label{formule.gammasnu.M}
2-\eta^M&=&1.657(5)\\
\label{formule.eta.M}
\eta^M&=&0.345(5)
\end{eqnarray}
The fit has been done discarding the two smallest sizes ($L=20$ and $L=40$)
which show small corrections. This value is different from the standard XY
case where $\eta=0.25$. Notice also that it   is in contradiction with the
result of Monte Carlo simulations
in the high temperature region obtained by Jos\'e {\it et al.} \cite{jose}
for the FFXY (which is believed to be in the same universality class as our
model) where $\eta\sim0.20$ was found. 
To our knowledge, it seems  one 
of the first times  this exponent 
is  calculated using finite size scaling.
From a theoretical point of view the $KT$ transition has  an 
exponential behavior, i.e.  a correlation length of the form 
$\xi\sim\exp[B_0\,(T-T_c)^{-\nu}]$,
however a power law behavior like
($\xi\sim(T-T_c)^{-\nu}$) can not be excluded {\it numerically}. In the
latter case  
the critical exponent $\nu$ can be 
calculated with the cumulant $V_2^M$ (\ref{formuleV2.M}). We have obtained
$\nu^M=0.92(3)$. With the finite size scaling method we were not 
able to compute  the 
exponent $\nu$ in the case of an exponential behavior. 

As for the Ising order parameter, the calculation of the exponents have been
done at the critical temperature but an error on $T_c^M$ leads to errors 
on the exponents, it is thus interesting to
find them without the help of $T^M_c$. This can be done using the same method 
as described before.
We have shown in \cite{LoisonO2Ferro} that this method is accurate enough
in order to obtain $\eta$ whatever the type of the behavior is (power law
or exponential). In Fig.~\ref{X2.1.KT}-\ref{X2.3.KT} we show our results
for three values of $\eta^M$. Obviously the value $\eta^M=0.33$ is the
best and we are able to obtain:
\begin{eqnarray}
\label{formule.eta2.M}
\eta^M=0.33(2) 
\end{eqnarray}
which is compatible with (\ref{formule.eta.M}). Moreover this result is 
a strong indication that our choice of the critical temperature is correct.
Indeed another choice leads to other non-compatible exponents. For example,
had we chosen $T_c^M=T_c^\kappa=0.56465$ we would obtain 
$\eta^M(T=T_c)=0.47$ 
which is incompatible with (\ref{formule.eta2.M}).

To sum up, we have computed for the first time  the critical exponent
$\eta^M=0.345(5)$ for the Kosterlitz-Thouless transition using the finite
size scaling region in Monte Carlo simulations. We have given strong 
arguments that, in our range of sizes, the critical temperature
for this transition is less than the critical temperature corresponding to
the Ising-like transition ($T_c^M<T_c^\kappa$). Note that this is in agreement
with the phase diagram of the Ising-XY model (Fig \ref{fig.I-XY}), where
if the two transitions never merge, we have $T_{KT}<T_I$ (see also ref.
\cite{koslee}).  We cannot exclude, contrary
to the ferromagnetic XY model, a power law behavior at $T>T_c^M$ which
should be characterized besides the exponent $\eta^M$, by $\nu^M\sim0.92$ and 
$U^M_*\sim0.638$.

\section{Conclusion}
In this paper we have  investigated the critical behavior of the $2D$,
$J_1-J_2$ XY model on the square lattice. We have first theoretically argued
that this model should be in the same universality class as the Ising-XY model
for $2J_2>J_1$. We have then carried on extensive Monte Carlo simulations
for the particular ratio $J_2/J_1=0.7$. Our main conclusion is that this
system undergoes two 
distinct and separate transitions. The first one is of Kosterlitz-Thouless
type
with an exponent $\eta=0.345(5)$ different to the ferromagnetic case, 
whereas the second one (associated to the chirality order parameter) seems to 
be in a non-Ising universality class. The temperatures of transitions and 
the values of the critical exponents are summarized in table~\ref{table3}. 
It is worth mentioning that the estimate of the exponents can be 
obtained  without the help of a precise determination of the critical 
temperatures.
The values of the critical exponents associated to the Ising symmetry 
are consistent with those obtained in
various recent works for different frustrated XY models
\cite{knops,jose,slee,luo,koslee}.
Nevertheless, our analysis has been done at $J_2/J_1=0.7$. We expect that
the exponents we obtained could vary with the ratio $J_2/J_1$ which makes
accurate comparisons difficult. Consequently, numerically speaking, we
cannot safely state that the $J_1-J_2$ XY model is in the same universality
class as other models quoted above.

The fact that two transitions exist at two different
temperatures and that the critical exponents of the chiral order parameter 
transition is not of Ising type
seems puzzling. How could we reconciliate them ?
One first idea, is to use an argument by  Olsson \cite{olsson} which
explains these strong deviations from the Ising
universality class by a large screening length (due to vortices) which
prevents  observing
the expected  Ising behavior. 
In our case, we would therefore expect for large sizes, a crossover to
such  behavior, i.e. for example, $\nu$ grows
to reach the value 1 for an infinite size. However no sign of this crossover
has appeared for our largest sizes ($L=150$). Obviously, such a crossover
could not be excluded for much larger size ($L\sim 1000$) but  seems not
very plausible. Two more 
 plausible interpretations can be {\it a priori} formulated.
One, due to Granato et al. \cite{granato} consists in invoking a new
universality class for the chiral order parameter. The idea of the
$3-$state Potts model universality class has for example been recently
advocated in \cite{slee}. Another one is due to 
Knops {\it et al.} \cite{knops}. These authors have introduced a
possible instable fixed point on the critical line PT of the phase diagram
of the Ising-XY model (see Fig. \ref{fig.I-XY})
which is able to induce strong cross over phenomena in the infrared limit.
This conjecture has the advantage to explain the whole set of dispersed
results found in the literature.
Notably, it would explain the continuous variation of exponents found by Granato {\it et al.} in the Ising-XY model \cite{isxy} but also the the $J_2/J_1$ dependence
of critical exponents in the $J_1-J_2$ XY model under consideration here.
 To answer these questions, very high precision Monte-Carlo simulations
for large size systems
could bring some answers to this problem. 
Moreover new analytical developments are deeply needed.

\section {Acknowledgments}
This work was supported by the Alexander von Humboldt Foundation and 
the EC TMR Program {\it Integrability, non-perturbative effects and 
symmetry in Quantum Field Theories}, grant FMRX-CT96-0012

\newpage
\begin{table}[t]
\begin{center}
\begin{tabular}{c|c|c}
\hspace{-10pt}
\begin{tabular}{c}$NOR$ \end{tabular}
\hspace{-10pt}
&
\begin{tabular}{c}$\tau_\kappa$\end{tabular}
\hspace{-10pt}
&
\begin{tabular}{c}$\tau_\kappa\, (1+NOR/6)$\end{tabular}
\hspace{-10pt}
\\
\hline
0&256(9)&256(9)
\hspace{-10pt}
\\
\hline
2&18.9(4)&25.1(5)
\hspace{-10pt}
\\
\hline
3&14.5(2)&21.7(3)
\hspace{-10pt}
\\
\hline
4&12.3(2)&20.5(3)
\hspace{-10pt}
\\
\hline
5&11.2(1)&20.5(2)
\hspace{-10pt}
\\
\hline
6&10.2(2)&20.3(3)
\hspace{-10pt}
\\
\hline
7&9.5(2)&20.5(3)
\hspace{-10pt}
\\
\hline
10&8.4(1)&22.4(2)
\hspace{-10pt}
\\
\hline
15&7.50(4)&26.2(1)
\hspace{-10pt}
\end{tabular}
\end{center}
\caption{\protect
\label{table1}
Autocorrelation times for the chirality for $L=30$ in function of the number
of over-relaxation steps $NOR$.
}
\end{table}

\newpage

\begin{table}[t]
\begin{center}
\begin{tabular}{c|c|c|c|c}
\hspace{-10pt}
\begin{tabular}{c}$L$ \end{tabular}
\hspace{-10pt}
&
\begin{tabular}{c}$t_{up}$\end{tabular}
\hspace{-10pt}
&
\begin{tabular}{c}$t_{av}$\end{tabular}
\hspace{-10pt}
&
\begin{tabular}{c}$\tau_\kappa$\end{tabular}
\hspace{-10pt}
&
\begin{tabular}{c}$t_{av}/\tau_\kappa$\end{tabular}
\hspace{-10pt}
\\
\hline
20& $5. 10^5$ & $20. 10^6$&7.95(13)&$2.5\,10^6$
\hspace{-10pt}
\\
\hline
40& $5. 10^5$ & $15. 10^6$&13.19(6)&$1 \,10^6$
\hspace{-10pt}
\\
\hline
60& $7. 10^5$ & $18. 10^6$&19.17(17)&$9 \,10^5$
\hspace{-10pt}
\\
\hline
80& $8. 10^5$ & $18. 10^6$&25.58(33)&$7\,10^5$
\hspace{-10pt}
\\
\hline
100& $1. 10^6$ & $16. 10^6$&31.66(50)&$5\,10^5$
\hspace{-10pt}
\\
\hline
120& $2. 10^6$ & $20. 10^6$&40.6(10)&$5\,10^5$
\hspace{-10pt}
\\
\hline
150& $3. 10^6$ & $32. 10^6$&50.9(15)&$6\,10^5$
\hspace{-10pt}
\end{tabular}
\end{center}
\caption{\protect
\label{table2}
Number of Monte Carlo steps to thermalize $T_{up}$ and to average 
$T_{av}$ as function of the size of the lattice $L$.
$\tau_\kappa$ are calculated with shorter MC runs. 
The last column gives the number of ''independent'' measures
which are, at least, one or two orders greater than previous studies. 
}
\end{table}

\newpage
\begin{table}[t]
\begin{center}
\begin{tabular}{c|c|c|c|c|c|c}
\hspace{-10pt}
\begin{tabular}{c}symmetry \end{tabular}
\hspace{-10pt}
&
\begin{tabular}{c}$T_c$ \end{tabular}
\hspace{-10pt}
&
\begin{tabular}{c}$U_*$ \end{tabular}
\hspace{-10pt}
&
\begin{tabular}{c}$\nu$ \end{tabular}
\hspace{-10pt}
&
\begin{tabular}{c}$\gamma/\nu$ \end{tabular}
\hspace{-10pt}
&
\begin{tabular}{c}$\eta$ \end{tabular}
\hspace{-10pt}
&
\begin{tabular}{c}$\beta/\nu$\end{tabular}
\hspace{-10pt}
\\
\hline
$Z_2$&0.56465(8)&0.6269(7)&0.795(20)&1.750(10)&0.250(10)$^1$&0.127(10)
\hspace{-10pt}
\\
\hline
$U(1)$&0.56271(5)&0.638(5)&0.92(3)$^2$& &0.345(5)&
\hspace{-10pt}
\end{tabular}
\end{center}
\caption{\protect\label{table3}
Summary of our results for the Ising symmetry ($Z_2$) and the
$XY$ symmetry $U(1)$. $^1$calculated using $2-\eta=\gamma/\nu$.
$^2$for a power law behavior.  
}
\end{table}

\newpage

\appendix

\section{Calculation of errors }
\label{annexe1}
We describe here our procedure to calculate statistical errors for the 
different quantities. 
The first stage is to find the number of independent steps in our Monte Carlo.
Indeed the Monte Carlo is a Markov process and therefore two consecutive
steps are correlated. 

\subsection{Autocorrelation time}
We define the autocorrelation function:
\begin{eqnarray}
\label{equrho}
\rho(t)={<A(0)A(t)>-<A(t)>^2 \over <A(t)^2>-<A(t)>^2},
\end{eqnarray}
where $A(t)$ is a thermodynamic quantity ($\kappa$,
$\chi^M$, $\chi^\kappa$ ...).
An example is shown in figure \ref{figurerho} for the chirality 
$\kappa$ at the temperature
$T=T_c^\kappa$ and for a lattice size $L=20$. 
$N_{MC}=50$ millions steps of Metropolis algorithm are used after
discarding 1 millions steps to thermalize the system.
We  calculate the autocorrelation time following the procedure 
of \cite{Madras88} by $\tau=\sum_{t=0}^{t_f} \rho(t)$ 
where $t_f$ is calculated in a self consistent way as $\rho(t)<0.01$
which corresponds to $t_f\sim5\tau$. In this case the  value we get 
is systematically underestimated for less than $1$ percent. 
It is important to 
stop the summation because the variance of $\tau$ is of the
order of the number of summation ($t_f$) and thus the error proportional 
to 
$\sqrt{t_f}$. 
Madras and Sokal \cite{Madras88} have proposed an estimation
of the error:
\begin{eqnarray}
\Delta\tau &\sim& \sqrt{2(2\,t_f+1)\tau^2/N_{MC}} \, .
\end{eqnarray}
However this formula seems to underestimate the result. Indeed we   calculate 
$\tau(\Delta\tau)$ function of the number of MC steps. We  obtain 
104(5), 112(2) and 116.2(8) for one, ten and fifty millions respectively.
Obviously the errors are underestimated ($104+5 < 116.2-0.8$). 
Therefore, 
in order to compute the errors for $\tau$ we make several simulations
for different
initial configurations and use the results as independent quantities to 
calculate the average and estimate the error. 

\subsection{Statistical Errors}
The second step after having computed the autocorrelation time $\tau$, is to 
calculate the error on each quantities. As they depend only on a
single average like the chirality or the 
susceptibility $\chi^K_2$ (\ref{formuleXK2}), the result is straightforward 
\cite{MullerB73}: 
\begin{eqnarray}
(\Delta<\kappa>)^2&=&(<\kappa^2>-<\kappa>^2){(1+2\,\tau/\tau_s) 
\over N_{MC}/\tau_s}, \\
(\Delta\chi_2)^2&=&(<\kappa^4>-<\kappa^2>^2){(1+2\,\tau/\tau_s) 
\over N_{MC}/\tau_s} {N \over k_B T},
\end{eqnarray}
where $N_{MC}$ is the number of Monte Carlo steps to average, 
$\tau_s$ the number of steps between two measurements
and $N$ the number of lattice sites.
Choosing $\tau_s=1$ we obtain $(1+2\,\tau)/N_{MC}\sim 2\,\tau/N_{MC}$ for 
large $\tau$ while  choosing $\tau_s=\tau$ we get $3\,\tau/N_{MC}$
which gives larger errors.

Problems arise when quantities are a combination of different averages,
for example the chirality $\chi^\kappa$ (\ref{formuleXK}). 
We could try to treat $<\kappa>$ and $<\kappa^2>$ 
as independent quantities and estimate
the error by the sum of the errors of the two quantities. However the
result will be overestimated due
to the correlation between the two elements of the sum.  
To solve this problem we can use, for example, the Jackknife procedure.
We do not review this method here but just present the essential points
we need (for a more complete review see \cite{Jackknife}).

In order to avoid being too abstract, we show how this method works
for the susceptibility
of the chirality (\ref{formuleXK}). For clarity we choose $\tau_s=1$. The
application for different $\tau_s$ is then straightforward. 
We have to define  
\begin{eqnarray}
\bar{\kappa}_t&=&{N_{MC} \, <\kappa>-\kappa_t \over N_{MC}-1}\\
\bar{\kappa^2}_t&=&{N_{MC} \, <\kappa^2>-\kappa^2_t \over N_{MC}-1}\\
\bar{\chi}_t&=&(\bar{\kappa^2}_t-\bar{\kappa}_t^2) {N \over k_B T}
\end{eqnarray}
where $t$ designs the MC step  and  $N_{MC}$ the total number of 
MC steps. Our estimate for the susceptibility and the
error will be given by:
\begin{eqnarray}
\label{formuleXjack}
\chi &\sim& {1 \over N_{MC}} \sum_{t=1}^{N_{MC}} \bar{\chi}_t \\
\label{formuleSigmajack}
\Delta\chi^2 &\sim& {N_{MC}-1 \over N_{MC}} \sum_{t=1}^{N_{MC}} 
(\bar{\chi}_t-\chi)^2*(1+2\,\tau) 
\end{eqnarray}
If we save the chirality at each MC step  the formulas are not 
difficult to apply. However we need a large hard disc to store the data.
For example if we wanted to save the 32 millions steps 
for the simulation of the
size $L=150$ we would then need 72 bytes for each step to save the energy, the 
magnetization and the chirality, which implies  more than two giga bytes.
To avoid
this problem, we could only save the data every $\tau_s=\tau$ but then the 
size of  the file would still be more than ten millions of bytes. 
Moreover we would lose informations and therefore errors would be greater.
We propose now a way which allows to obtain a good estimate without the 
problem of large storage and with minor changes in the program. 

We use a development for large $n=N_{MC}/(1+2\,\tau)$  
(which is always the case in MC), choosing $\tau_s=1$. In this 
case the formula (\ref{formuleXjack}-\ref{formuleSigmajack}) becomes, keeping
only the dominant term:
\begin{eqnarray}
\chi &\sim& (<\kappa^2>-<\kappa>^2) {N \over k_B T}\\
\label{formuleSigmajack2}
\Delta\chi^2 &\sim& {(1+2\,\tau) \over N_{MC}} \large{[}<\kappa^4>
-<\kappa^2>^2+4<\kappa>^2(<\kappa^2>-<\kappa>^2) \nonumber \\
&&-4 <\kappa>(<\kappa^3>-<\kappa^2><\kappa>)\large{]} ({N \over k_B T})^2
\end{eqnarray}
The chirality conserves its initial form while the error is the sum of the two
errors (of $<\kappa^2>$ and $<\kappa>^2$) subtracted by the third 
term which representing
the correlation between them. We note that this procedure induces 
a small change in the program: we have only to save in the histogram 
$<\kappa^3>$
plus the values of $<\kappa>$ and $<\kappa^2>$. 
To test our formula we have 
computed the susceptibility associated to the chirality (\ref{formuleXK}) and 
its errors calculated by three ways.
We perform the simulation with 
a lattice size $L=20$ with 4 steps of over relaxation algorithm between
each Monte Carlo, for one million steps. In this case the autocorrelation
time is about 8 (see table~\ref{table2}). 
The first method consists in saving at each step the energy, the
magnetization and
the chirality, the second
in  saving the data only at each $\tau$ steps, while 
the third in    using the 
approximate formula (\ref{formuleSigmajack2}).
We obtain, respectively: $\chi^\kappa=$ 11.22(8), 11.23(10), 11.22(8).
The three methods give compatible results but the third one gives the
best estimate
with the smallest size of storage (some hundred thousands bytes). 

We give hereafter the results of our calculation for the binder parameters
(\ref{formuleUM}-\ref{formuleUK}) and the cumulant $V_1$ and $V_2$ 
(\ref{formuleV1}-\ref{formuleV2}):
\begin{eqnarray}
\Delta{U}^2 &\sim& {(1+2\,\tau) \over N_{MC}} \large{[}
4 <\kappa^4>^3-4 <\kappa^2><\kappa^4><\kappa^6>
+8<\kappa^2>^2<\kappa^8> \nonumber \\
&&- <\kappa^2>^2<\kappa^4>^2 \large{]}\\
\Delta{V_1}^2 &\sim& {(1+2\,\tau) \over N_{MC}} \large{[}
{<\kappa E>^2 \over <\kappa>^2} ({<\kappa^2 E^2> \over <\kappa E>^2} 
- 2 {<\kappa^2 E>\over <\kappa E><\kappa>} + {<\kappa^2> \over<\kappa>^2} )
\nonumber \\
&&+ <E^2>-<E>^2 \nonumber \\
&&-2 {<\kappa E>\over <\kappa>} ({<\kappa E^2>\over<\kappa E>} 
- {<\kappa E>\over <\kappa>}) 
\large{]}\\
\Delta{V_2}&=& \Delta{V_1}\  substituing \ \kappa \ by\  \kappa^2
\end{eqnarray}

\subsection{Systematic Errors }
In addition to statistical errors, we have to take care of systematic errors. 
There are essentially of two kinds:
one due to the correlation between the random 
number and one due to  the use of the histogram technic.

The first one appears when we use a bad random generator. In this case
the period
of the random numbers could be very small and could thus introduce
correlations
between data. One example is the linear congruential generator 
used by many physicist for Monte Carlo simulations!
For certain choices of the initial parameter, the period could be very small
(less than 2000) and therefore could induce systematic errors. 
We use in this work a random generator with a period of more than 100 millions
found in Numerical Recipes (ran1) \cite{NumericalR}.

A second source of systematic error comes from the histogram technic
and the difference between the temperature of simulation and the temperature 
where the quantities are computed \cite{FerrenbergLS95,Newman98}.
In our simulation this difference is kept less than 0.005 in order to minimize
this effect.

\newpage
\begin{center}
FIGURE CAPTIONS
\end{center}

\begin{figure} 
\hspace{3cm}
\psfig{figure=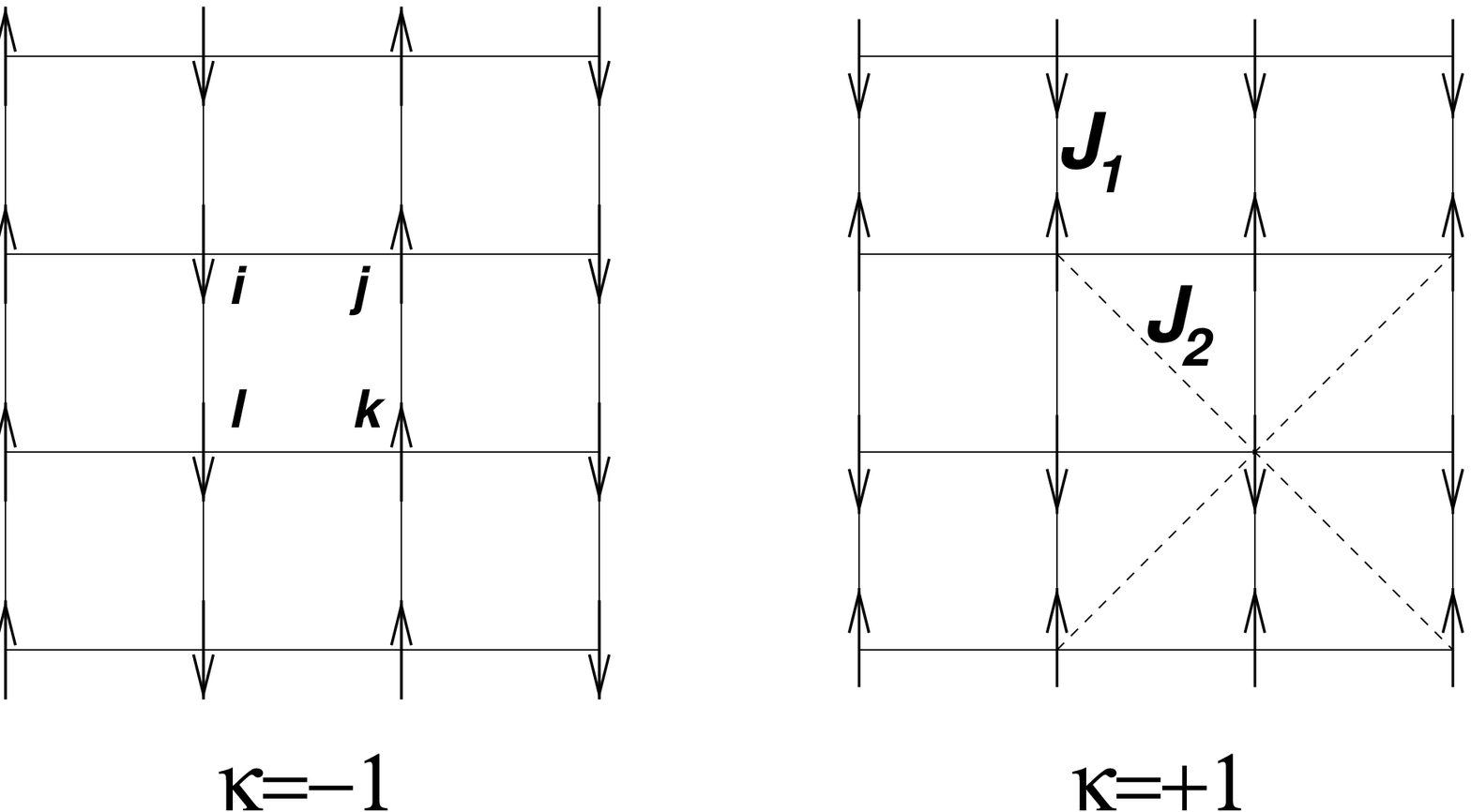,height=5cm}
\caption{}
\label{gs}
\end{figure}

Fig.~\ref{gs}:
Ground state of the $J_1-J_2$ XY model for $2J_2>J_1$

\begin{figure} 
\hspace{3cm}
\psfig{figure=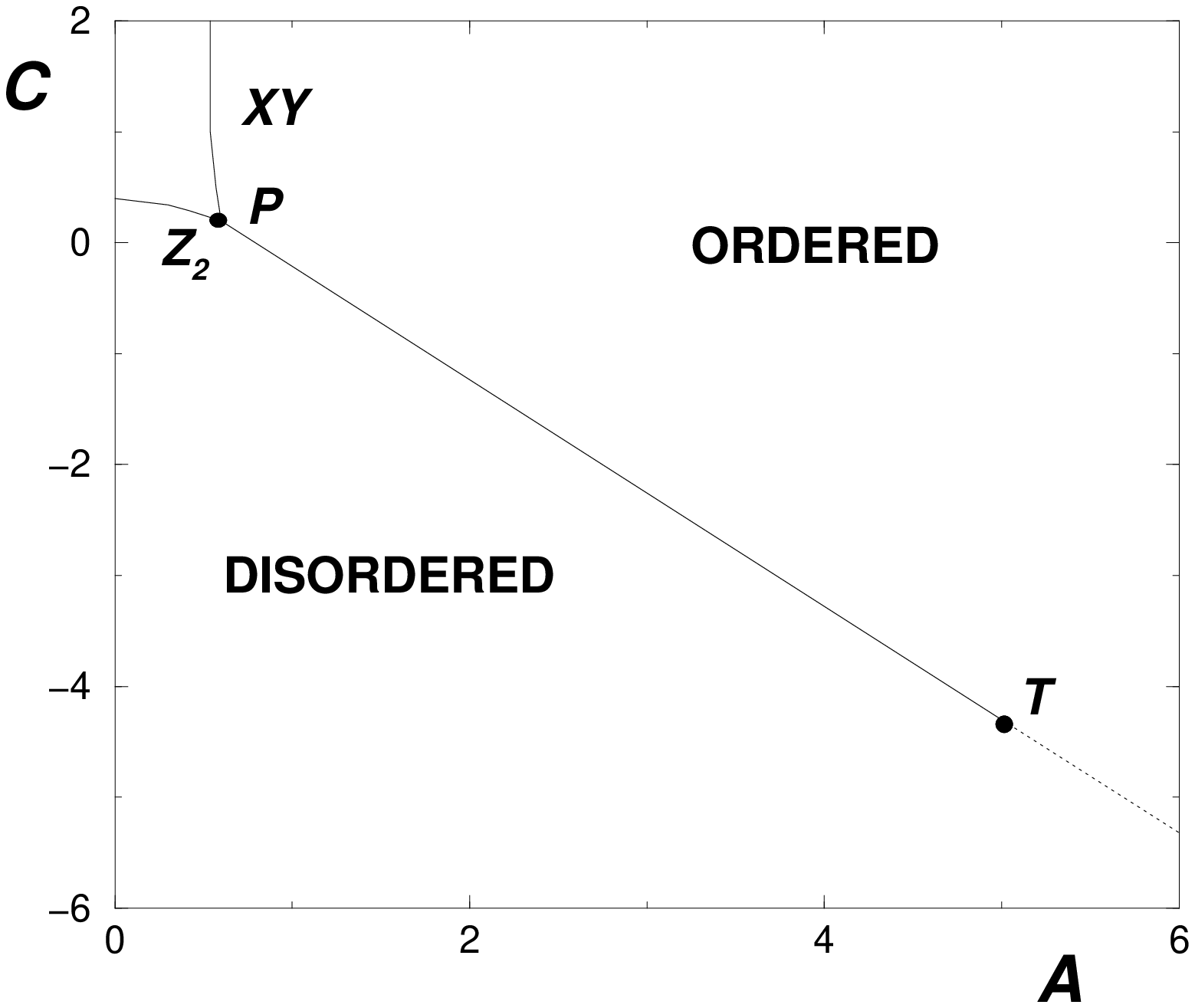,height=5cm}
\caption{
}
\label{fig.I-XY}
\end{figure}

Fig.~\ref{fig.I-XY}:
Phase diagram of the Ising-XY model.
Solid and dotted lines indicate continuous and first-order
transitions respectively.

\begin{figure} 
\hspace{3cm}
\psfig{figure=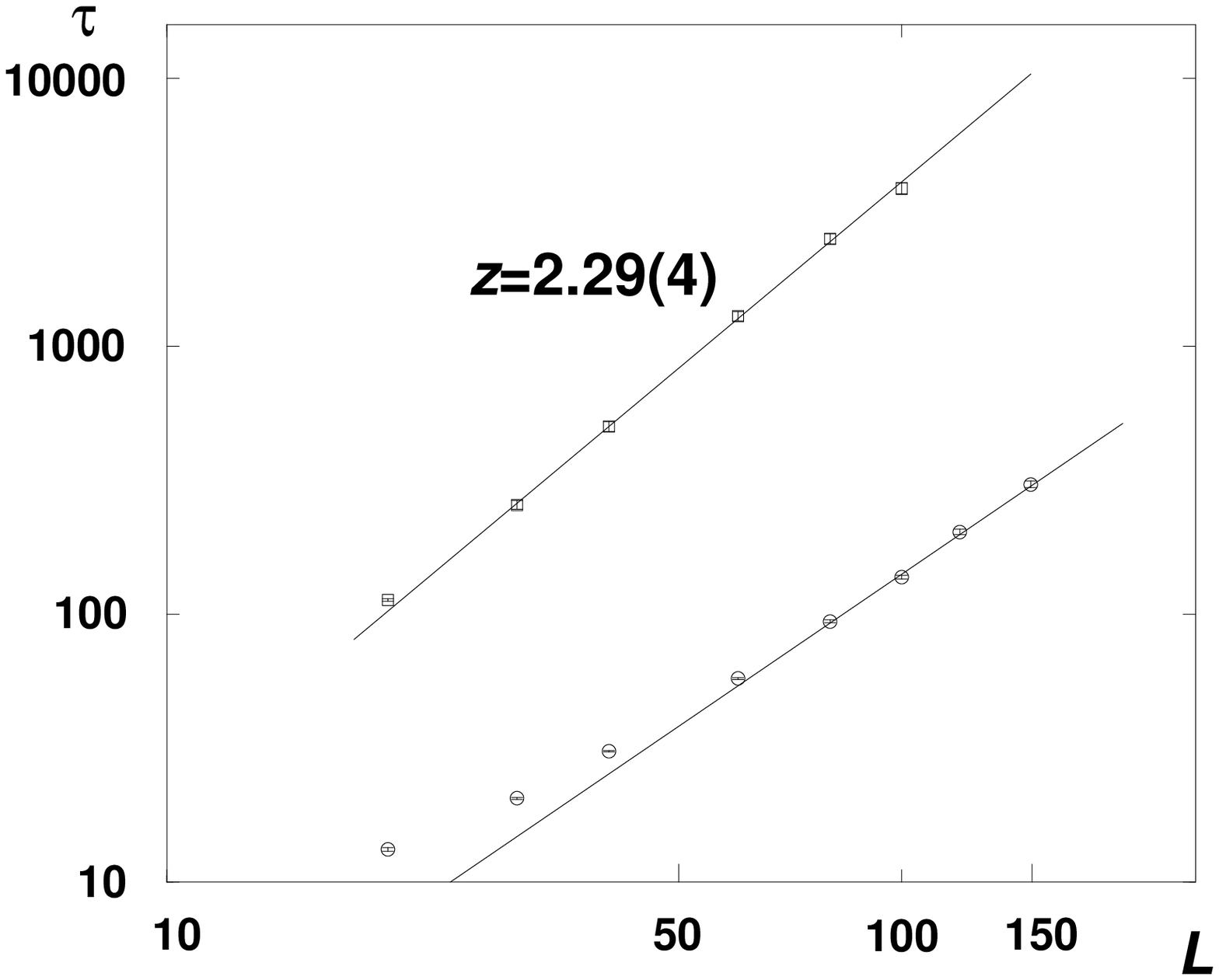,width=8.5cm}
\caption{
}
\label{figuretau}
\end{figure}

Fig.~\ref{figuretau}:
Real autocorrelation time for the standard Metropolis algorithm
(circle) and in combination with the over-relaxation algorithm (square).

\begin{figure} 
\hspace{3cm}
\psfig{figure=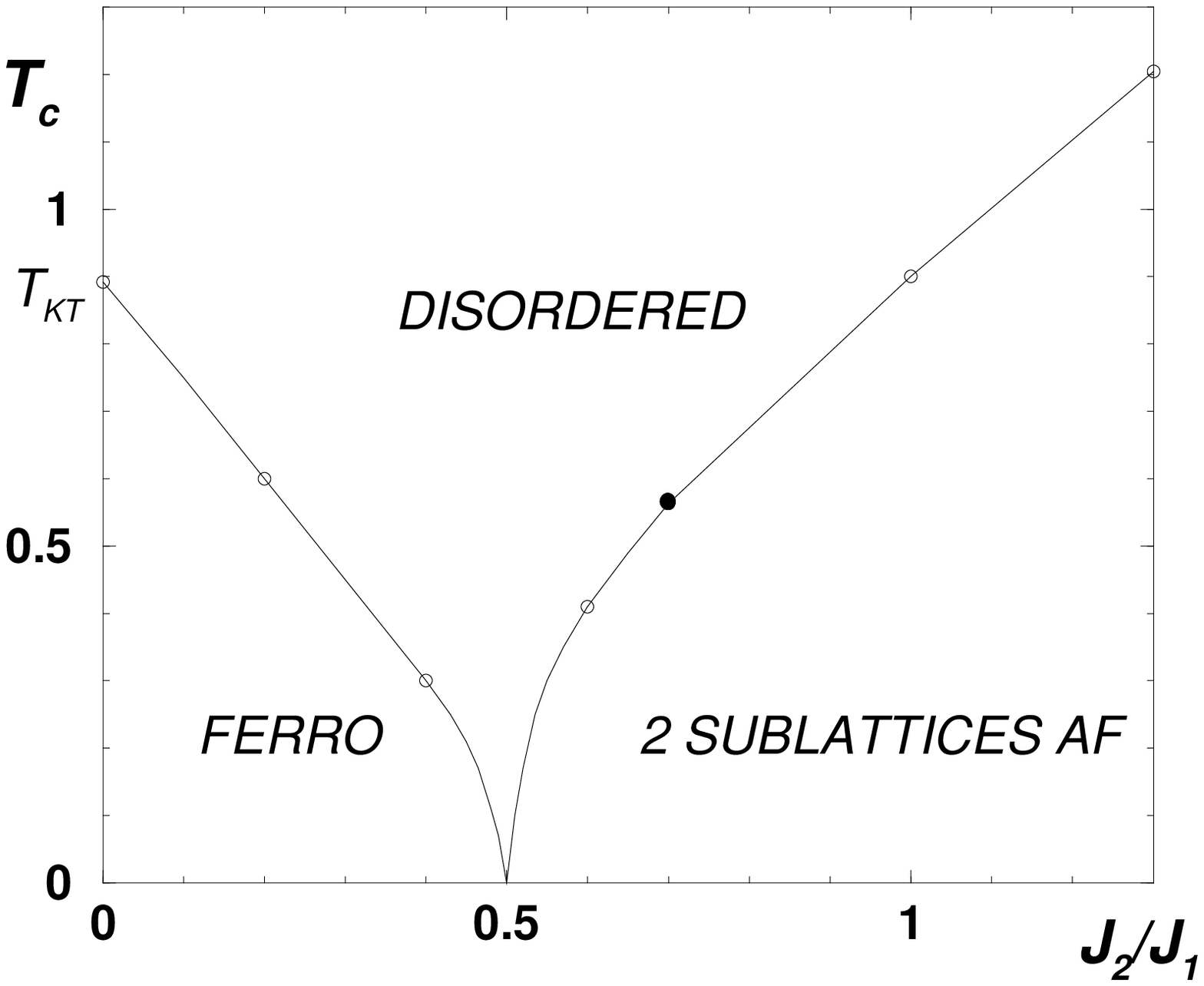,width=8.5cm}
\caption{
\label{figureTc}
}
\end{figure}

Fig.~\ref{figureTc}:
Phase diagram for the $J_1-J_2$ model. For $2J_2<J_1$ we find the normal
Kosterlitz-Thouless transition. Lines are just guides for the eyes.
Our study is done for $J_2/J_1=0.7$ (black circle).

\begin{figure} 
\hspace{3cm}
\psfig{figure=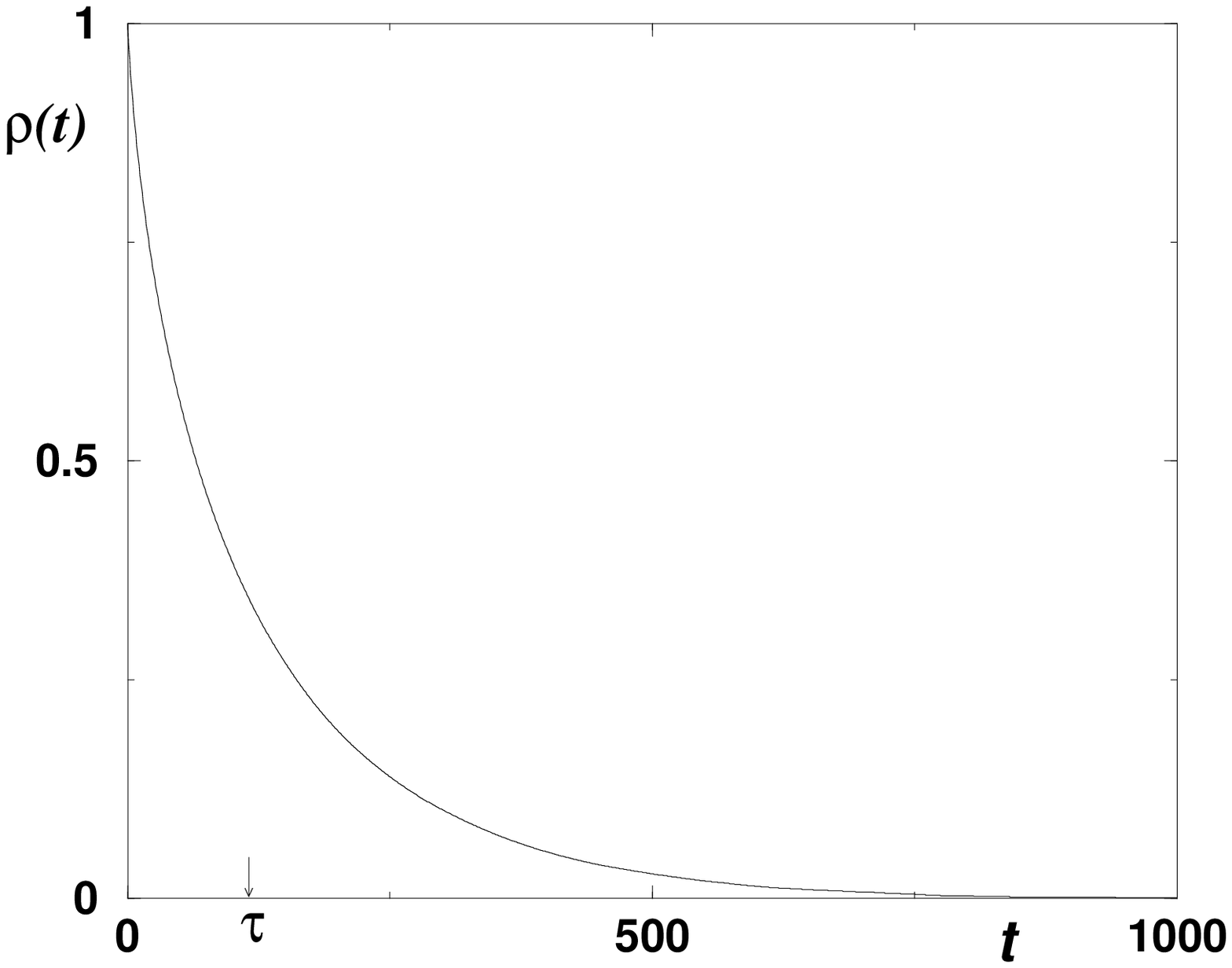,width=8.5cm}
\caption{
\label{figurerho}
}
\end{figure}

Fig.~\ref{figurerho}:
Autocorrelation $\rho(t)$ for the chirality $\kappa$ at $T=T_c^\kappa$.
The lattice size is $L=20$ and the number of MC is $N_{MC}=50$ millions.
The estimated $\tau=116(2)$ is shown by an arrow.

\begin{figure} 
\hspace{3cm}
\psfig{figure=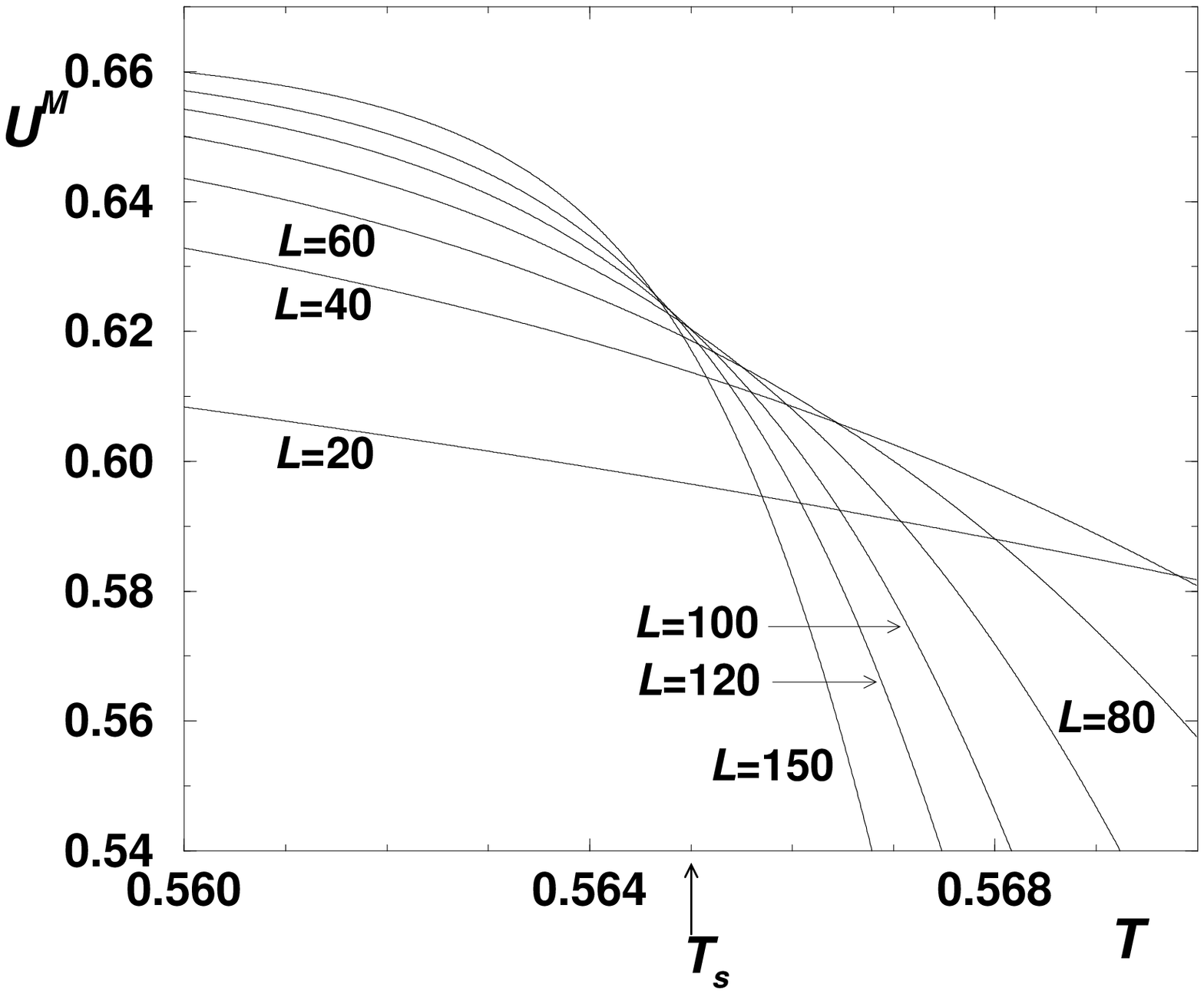,width=8.5cm}
\caption{
\label{figCM}
}
\end{figure}

Fig.~\ref{figCM}:
Binder's parameter $U^\kappa$ for the Ising order parameter
function of the temperature for various sizes $L$.
The arrow shows the temperature of simulation $T_s=0.565$.

\begin{figure} 
\hspace{3cm}
\psfig{figure=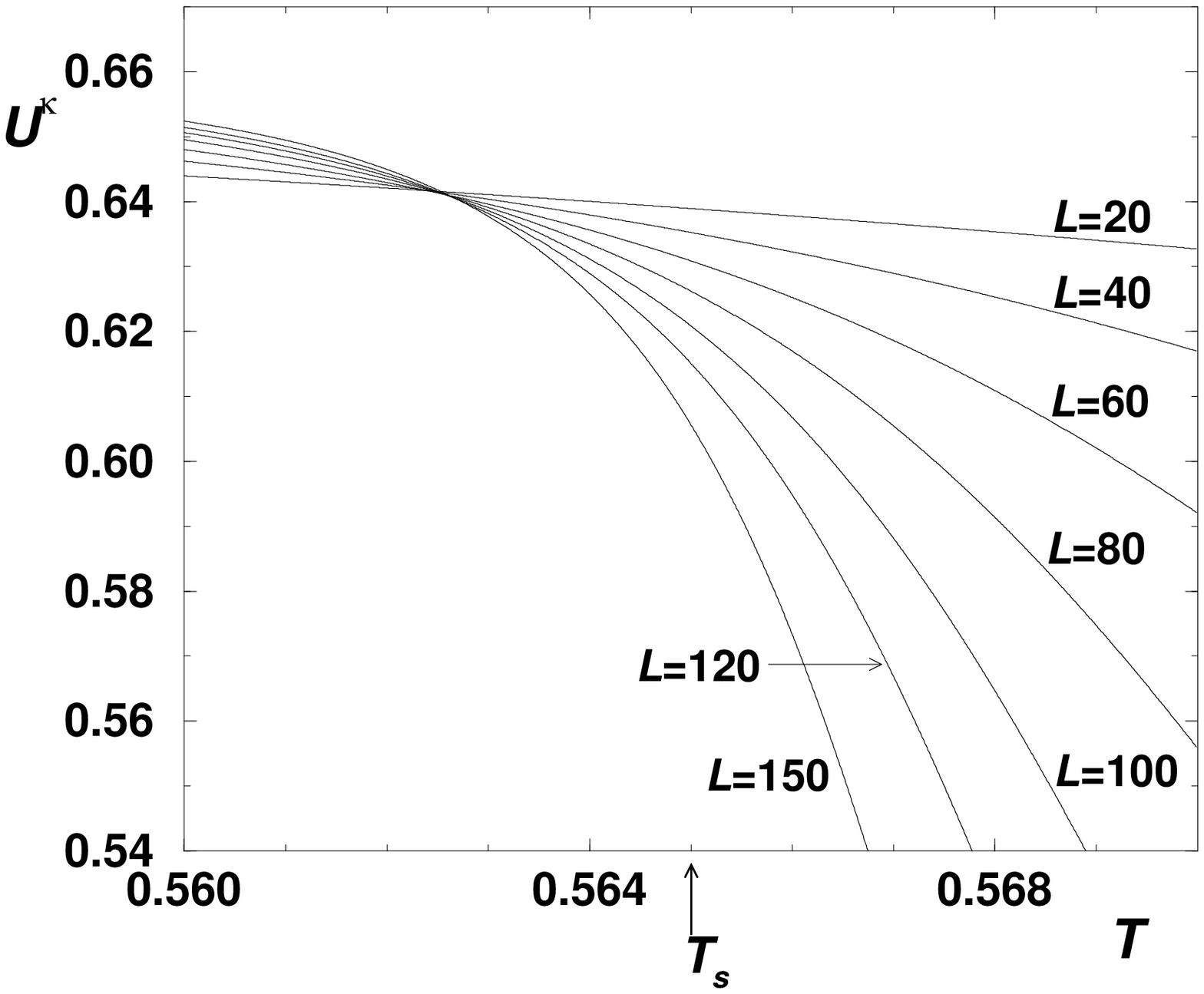,width=8.5cm}
\caption{
\label{figCM.KT}
}
\end{figure}

Fig.~\ref{figCM.KT}:
Binder's parameter $U^M$ for the $U(1)$ order parameter
function of the temperature for various sizes $L$.
The arrow shows the temperature of simulation $T_s=0.565$.
the scales is similar to those of Fig.~\ref{figCM}

\begin{figure} 
\hspace{3cm}
\psfig{figure=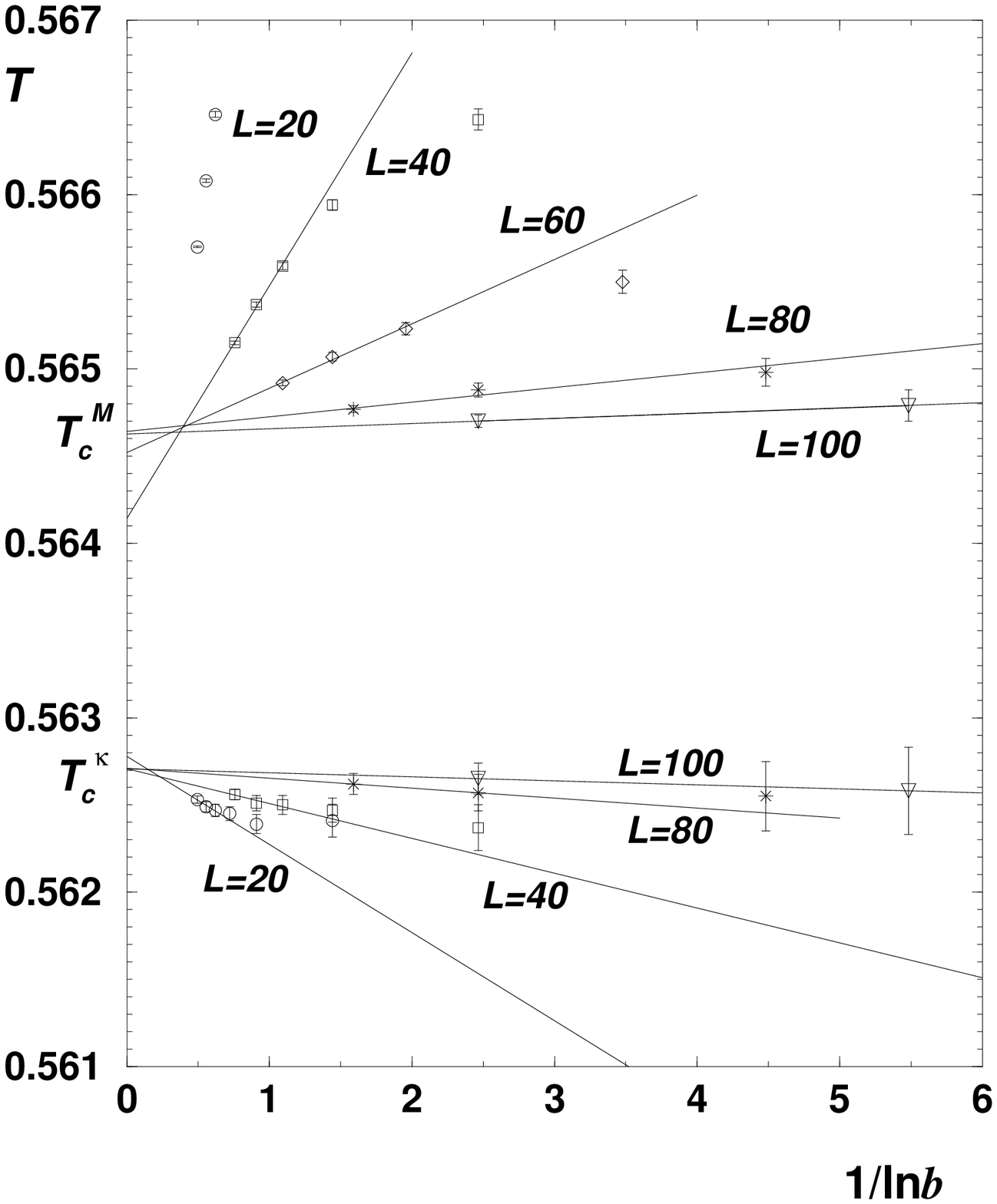,width=8.5cm}
\caption{
\label{figTc2.all}
}
\end{figure}

Fig.~\ref{figTc2.all}:
Crossing $T$ plotted vs inverse logarithm of the scale factor $b=L'/L$.
The upper part of the figure corresponds to $U^\kappa$ while the lower
part  to $U^M$. In the last case the size $L=60$
is not shown for clarity. We obtain $T_c^\kappa=0.56465(8)$ and
$T_{c}^M= 0.56271(5)$ with a linear fit (see text for comments).

\begin{figure} 
\hspace{3cm}
\psfig{figure=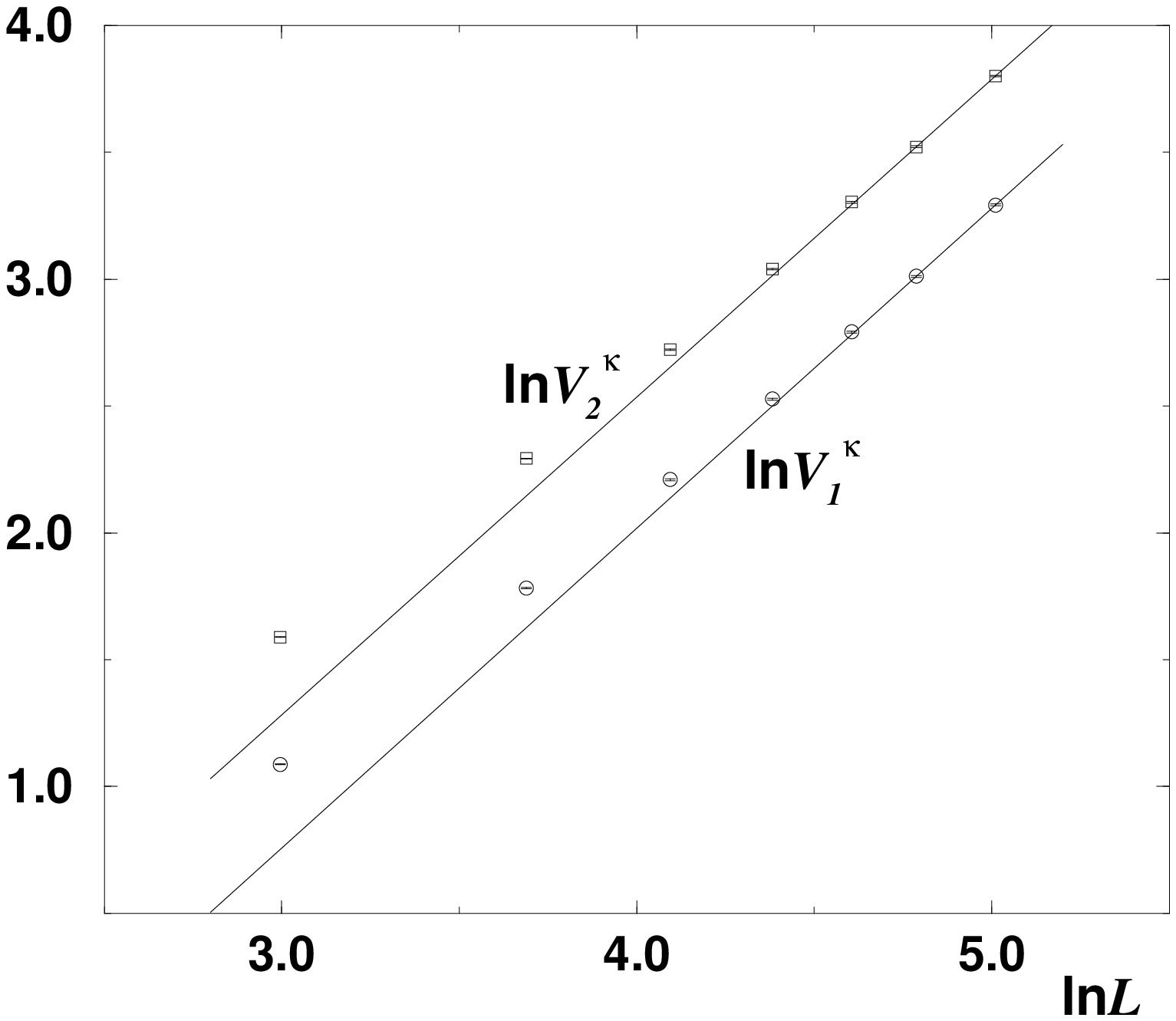,width=8.5cm}
\caption{
\label{figV}
}
\end{figure}

Fig.~\ref{figV}:
Values of $V_1^\kappa$ and $V_2^\kappa$  function of $L$ in a log--log
scale at $T_c^\kappa$. The value of the slopes gives $1/\nu^\kappa$. We observe
strong corrections for  small sizes. Only the three largest sizes
are used for the fits. When not shown, the estimated
statistical errors are smaller than the symbol.

\begin{figure} 
\hspace{3cm}
\psfig{figure=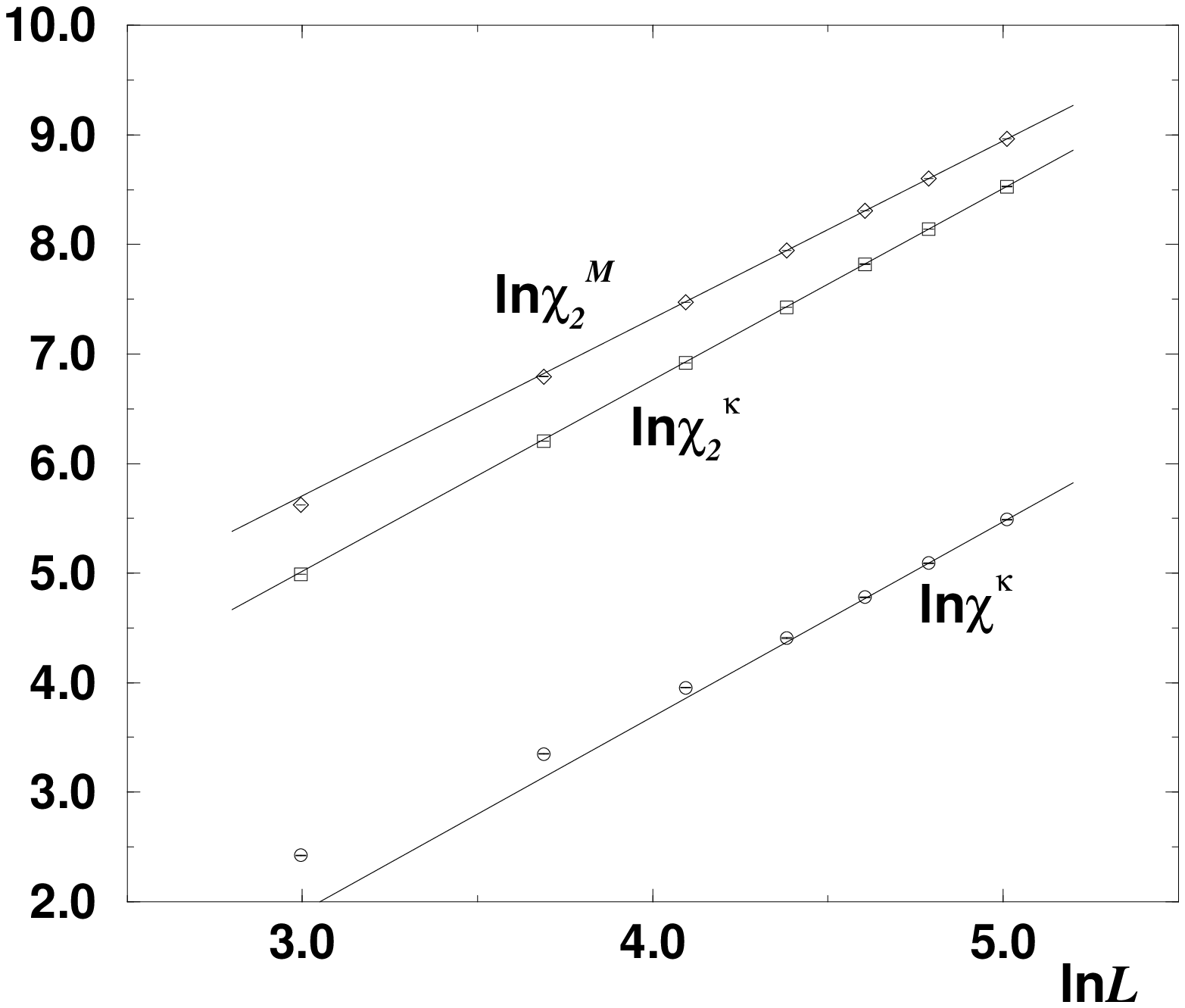,width=8.5cm}
\caption{
\label{figX.all}
}
\end{figure}

Fig.~\ref{figX.all}:
Values of $\chi^\kappa$ and $\chi_2^\kappa$  function of $L$ in a log--log
scale at $T_c^\kappa$ and $\chi_2^M$ at $T_c^M$.
The value of the slopes gives $\gamma/\nu=2-\eta$. We observe
strong corrections for the small sizes for $\chi^\kappa$. Only the three
largest sizes are used for the fit for $\chi^\kappa$ while only the smallest
sizes $L=20$ and $L=40$ are discarded for  the fits for $\chi_2^\kappa$
and $\chi_2^M$.
When not shown, the estimated
statistical errors are smaller than the symbol.

\begin{figure} 
\hspace{3cm}
\psfig{figure=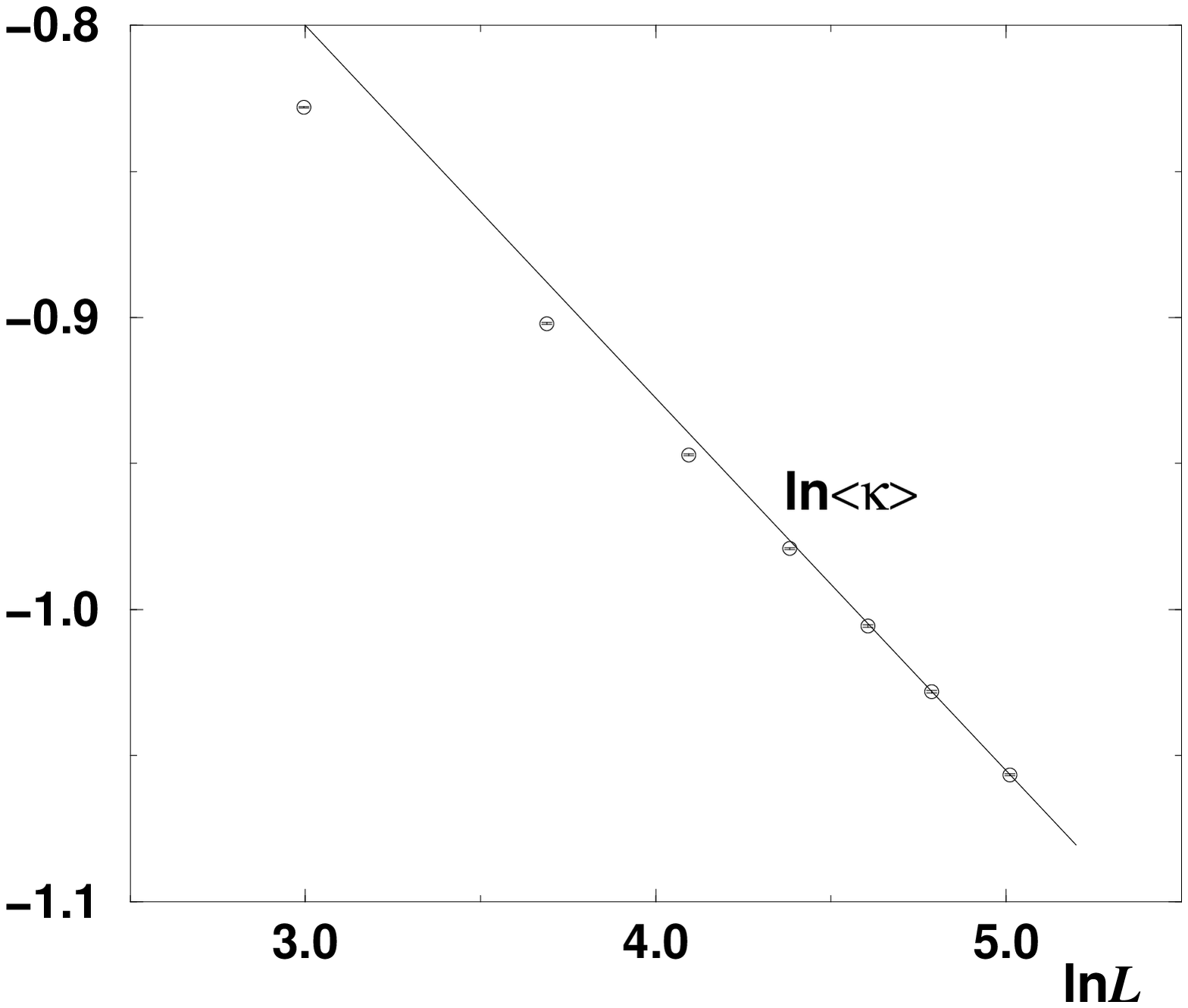,width=8.5cm}
\caption{
\label{figM}
}
\end{figure}

Fig.~\ref{figM}:
Values of $<\kappa>$ as function of $L$ in a log--log
scale at $T_c^\kappa$. The value of the slopes gives $\beta^\kappa/\nu^\kappa$.
We observe
strong corrections for the small sizes. Only the three largest sizes
are used for the fits. When not shown, the estimated
statistical errors are smaller than the symbol.

\begin{figure} 
\hspace{3cm}
\psfig{figure=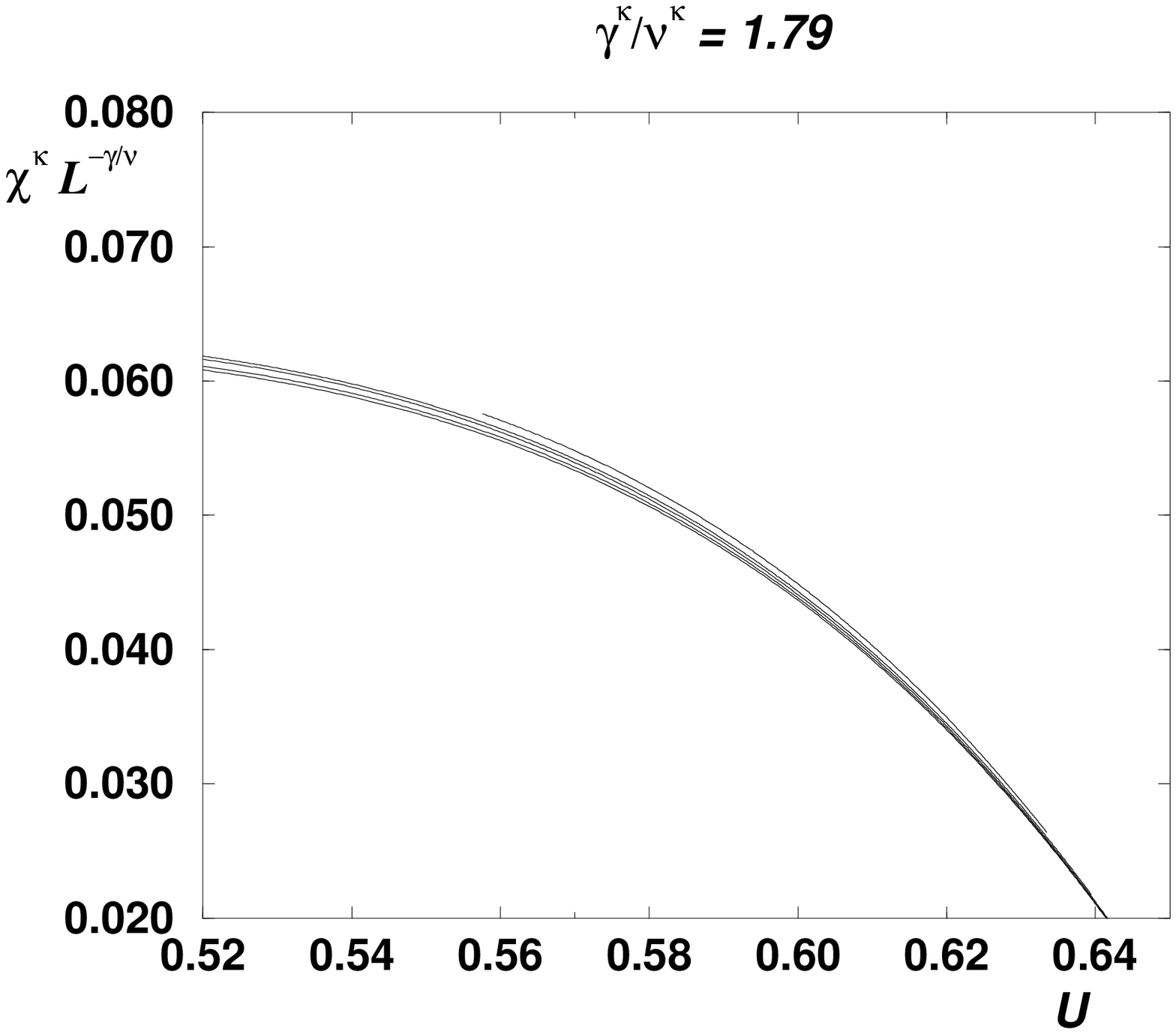,width=8.5cm}
\caption{
\label{X2.1}
}
\end{figure}

Fig.~\ref{X2.1}:
$\chi^\kappa L^{-\gamma^\kappa/\nu^\kappa}$  function of $U^\kappa$ with
$\gamma^\kappa/\nu^\kappa=1.79$ for the sizes $L=60,~80,~100,~120$ and 150.
The curves do not collapse in one curve.

\begin{figure} 
\hspace{3cm}
\psfig{figure=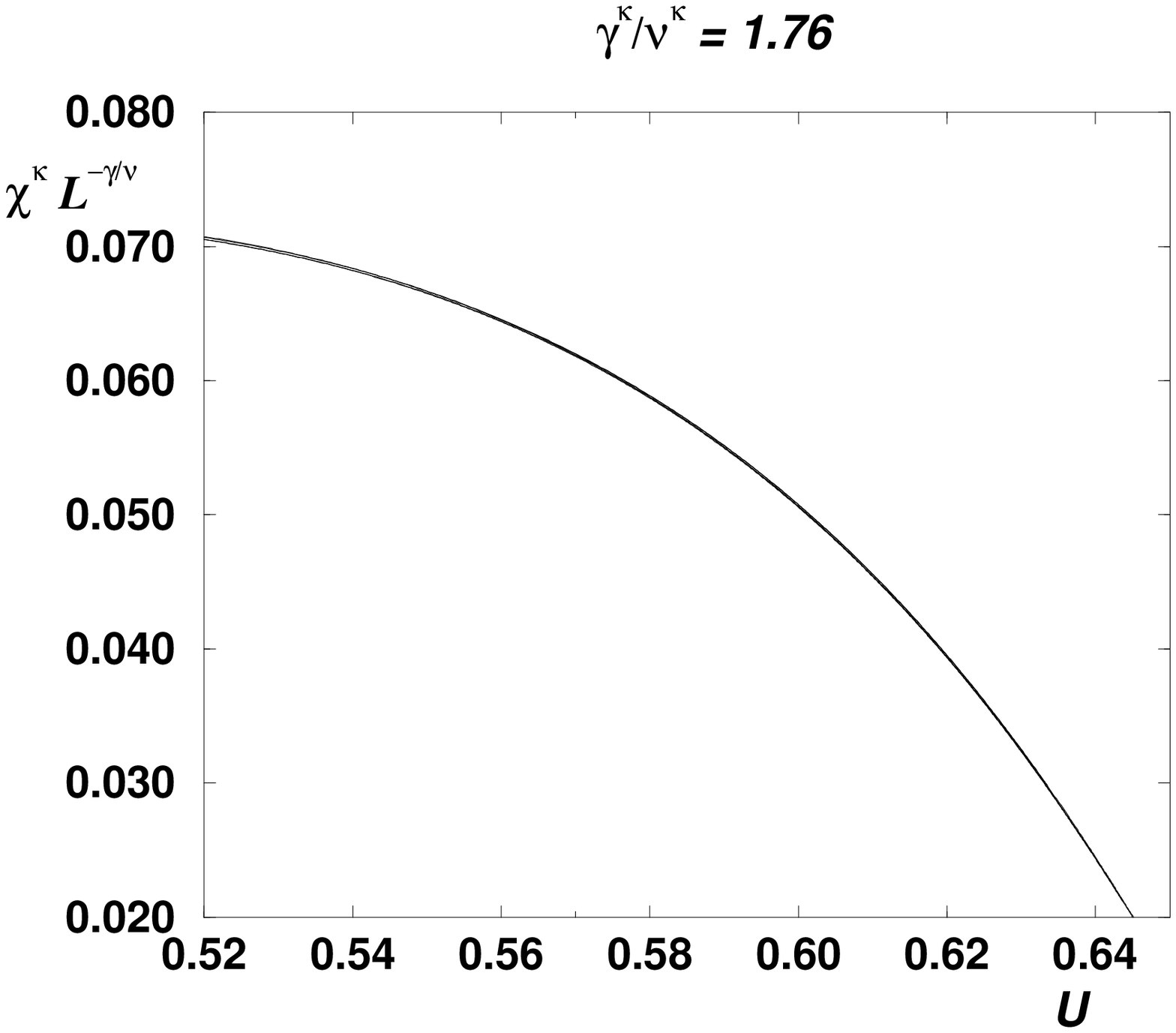,width=8.5cm}
\caption{
\label{X2.2}
}
\end{figure}

Fig.~\ref{X2.2}:
$\chi^\kappa L^{-\gamma^\kappa/\nu^\kappa}$ as function of $U^\kappa$ with
$\gamma^\kappa/\nu^\kappa=1.76$ for the sizes $L=60,~80,~100,~120$ and 150.
The curves collapse in one curve.

\begin{figure} 
\hspace{3cm}
\psfig{figure=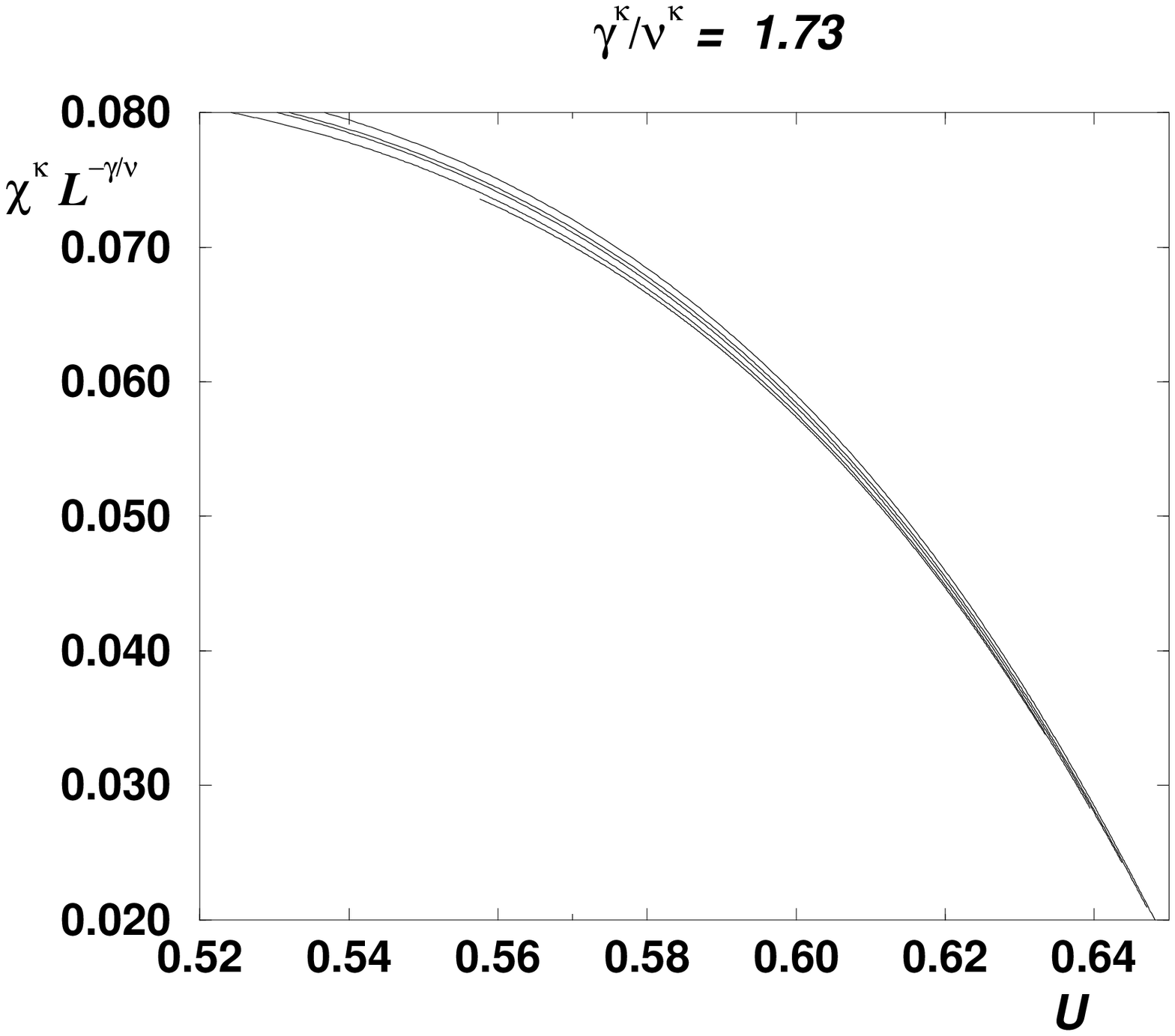,width=8.5cm}
\caption{
\label{X2.3}
}
\end{figure}

Fig.~\ref{X2.3}:
$\chi^\kappa L^{-\gamma^\kappa/\nu^\kappa}$  function of $U^\kappa$ with
$\gamma^\kappa/\nu^\kappa=1.73$ for the sizes $L=60,~80,~100,~120$ and 150.
The curves do not collapse in one curve.

\begin{figure} 
\hspace{3cm}
\psfig{figure=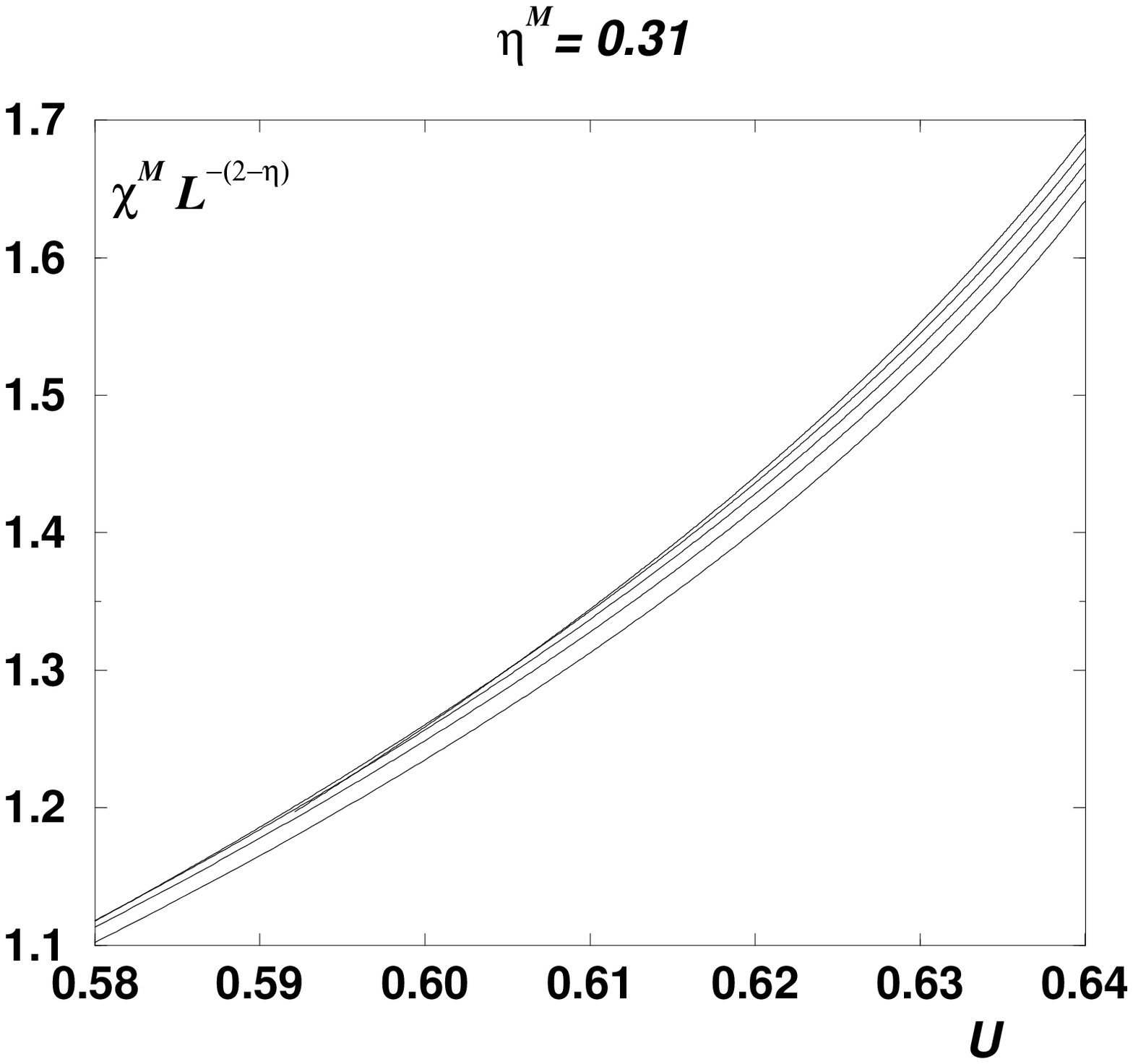,width=8.5cm}
\caption{
\label{X2.1.KT}
}
\end{figure}

Fig.~\ref{X2.1.KT}:
$\chi^M L^{2-\eta^M}$  function of $U^M$ with
$\eta^M=0.31$ for the sizes $L=60,~80,~100,~120$ and 150.
The curves do not collapse in one curve.

\begin{figure} 
\hspace{3cm}
\psfig{figure=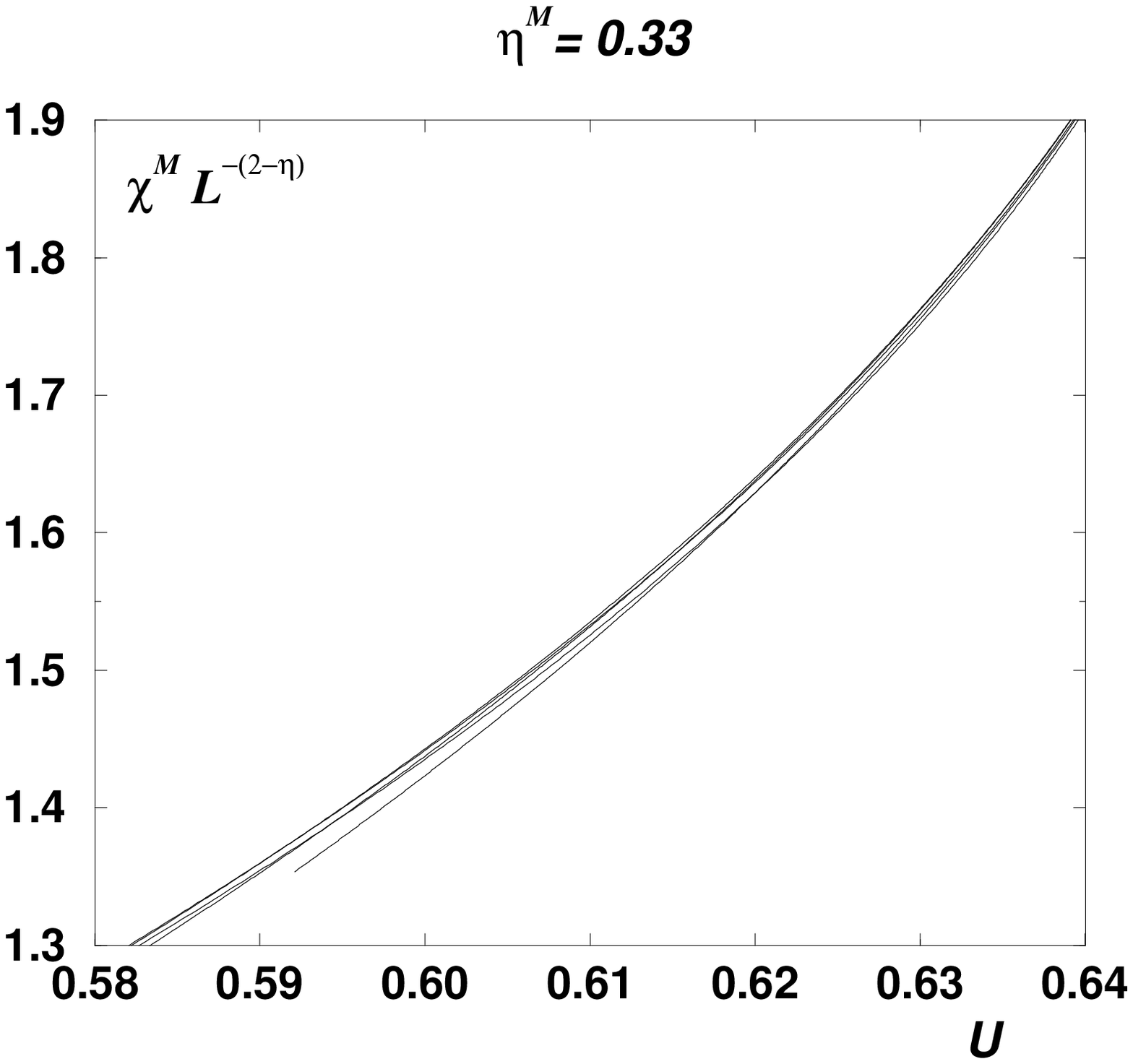,width=8.5cm}
\caption{
\label{X2.2.KT}
}
\end{figure}

Fig.~\ref{X2.2.KT}:
$\chi^M L^{2-\eta^M}$  function of $U^M$ with
$\eta^M=0.33$ for the sizes $L=60,~80,~100,~120$ and 150.
The curves collapse in one curve.

\begin{figure} 
\hspace{3cm}
\psfig{figure=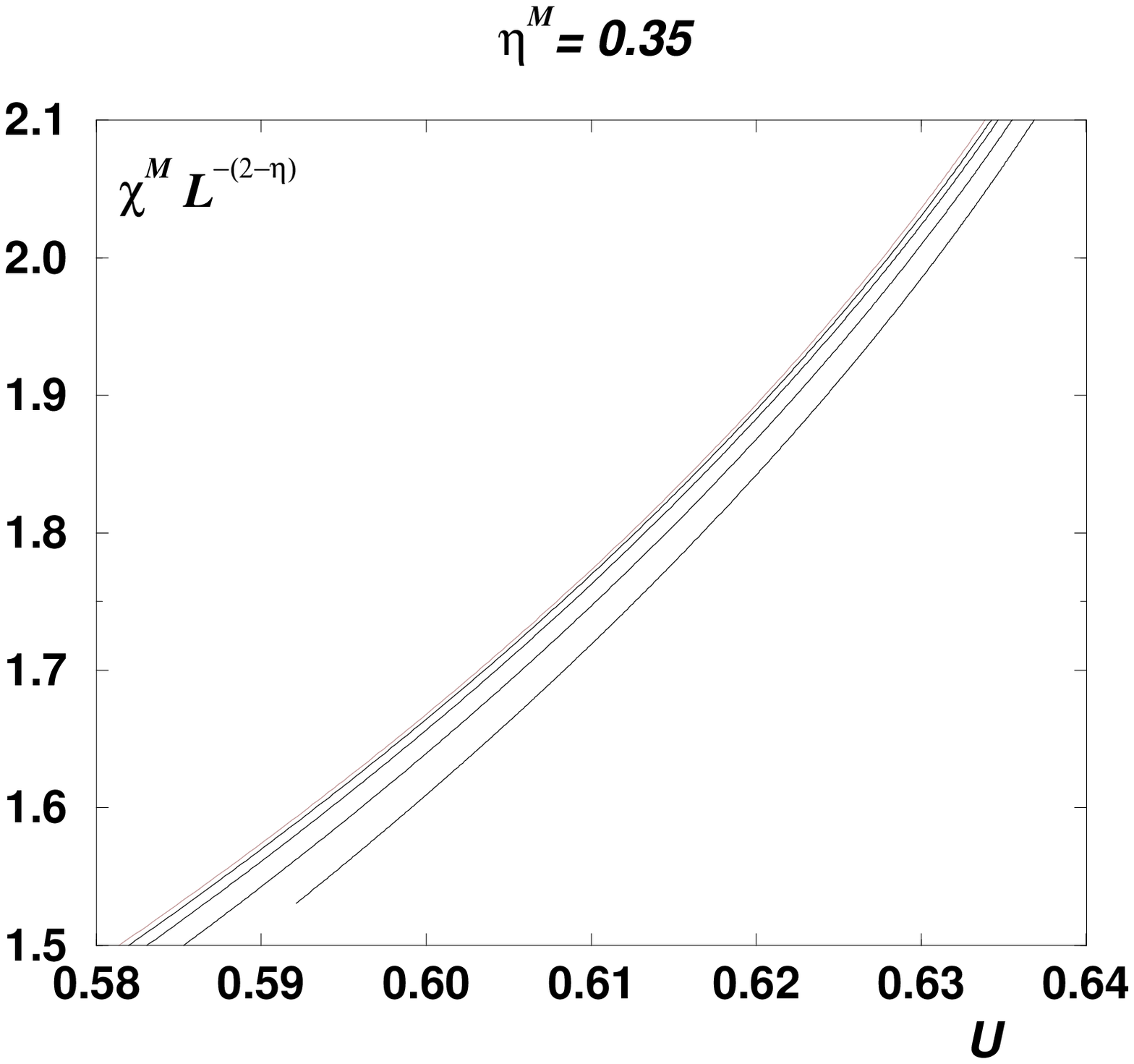,width=8.5cm}
\caption{
\label{X2.3.KT}
}
\end{figure}

Fig.~\ref{X2.3.KT}:
$\chi^M L^{2-\eta^M}$  function of $U^M$ with
$\eta^M=0.35$ for the sizes $L=60,~80,~100,~120$ and 150.
The curves do not collapse in one curve.

\newpage

\end{document}